\documentclass[twocolumn]{bmcart}

\usepackage{amsthm,amsmath,graphicx,float}
\RequirePackage{natbib}
\RequirePackage{hyperref}
\usepackage[utf8]{inputenc} 

\startlocaldefs
	\usepackage[version=3,arrows=font]{mhchem}
	\newcommand{\dx}{\dd x}
	\DeclareMathOperator{\Tr}{T}
	\DeclareMathOperator{\GR}{G}
	\newcommand{\Gdot}{\dot{\GR}}
	\newcommand{\Gddot}{\ddot{\GR}}
	\newcommand{\Ga}{\alpha}
	\newcommand{\Gb}{\beta}
	\newcommand{\Gd}{\delta}
	\newcommand{\Gf}{\phi}
	\newcommand{\Gg}{\gamma}
	
	\newcommand{\Gm}{\mu}
	\newcommand{\Gs}{\sigma}

	\newcommand{\dd}{\mathrm{d}}
	\newcommand{\ovr}[2]{{\frac{#1}{#2}}}
	\newcommand{\dovr}[2]{\ovr{\dd #1}{\dd #2}}

	\newcommand{\Eq}[1]{Eq.~(\ref{eq:#1})}

	\newcommand{\Fig}[1]{Fig.~\ref{fig:#1}}

	\newcommand{\Table}[1]{Table~\ref{box:#1}}

	\def\citeyear{\citep}
	\def\autocite{\citep}
\endlocaldefs

\begin{document}

\begin{frontmatter}

\begin{fmbox}
\dochead{Review}


\title{Input-output relations in biological systems: measurement, information and the Hill equation}


\author[
   addressref={aff1},                   
   corref={aff1},                       
   email={safrank@uci.edu}   
]{\inits{SA}\fnm{Steven A} \snm{Frank}}


\address[id=aff1]{
  \orgname{Department of Ecology and Evolutionary Biology}, 
  \street{University of California},                     %
  \city{Irvine, CA}                              
  \postcode{92697--2525}                                
  \cny{USA}                                    
}





\begin{abstractbox}

\begin{abstract} 
Biological systems produce outputs in response to variable inputs. Input-output relations tend to follow a few regular patterns.  For example, many chemical processes follow the S-shaped Hill equation relation between input concentrations and output concentrations.  That Hill equation pattern contradicts the fundamental Michaelis-Menten theory of enzyme kinetics. I use the discrepancy between the expected Michaelis-Menten process of enzyme kinetics and the widely observed Hill equation pattern of biological systems to explore the general properties of biological input-output relations.  I start with the various processes that could explain the discrepancy between basic chemistry and biological pattern. I then expand the analysis to consider broader aspects that shape biological input-output relations.  Key aspects include the input-output processing by component subsystems and how those components combine to determine the system's overall input-output relations.  That aggregate structure often imposes strong regularity on underlying disorder.  Aggregation imposes order by dissipating information as it flows through the components of a system.  The dissipation of information may be evaluated by the analysis of measurement and precision, explaining why certain common scaling patterns arise so frequently in input-output relations.  I discuss how aggregation, measurement and scale provide a framework for understanding the relations between pattern and process.  The regularity imposed by those broader structural aspects sets the contours of variation in biology.  Thus, biological design will also tend to follow those contours.  Natural selection may act primarily to modulate system properties within those broad constraints.
\parttitle{Published version} Frank, S.~A.\ 2013.\ Input-output relations in biological systems: measurement, information and the Hill equation.\ Biology Direct, \textbf{8}:31, \href{http://dx.doi.org/10.1186/1745-6150-8-31}{doi:10.1186/1745-6150-8-31}
\end{abstract}




\end{abstractbox}
\end{fmbox}

\end{frontmatter}




\section*{Introduction}
Cellular receptors and sensory systems measure input signals.  Responses flow through a series of downstream processes. Final output expresses physiological or behavioral phenotype in response to the initial inputs. A system's overall input-output pattern summarizes its biological characteristics. 

Each processing step in a cascade may ultimately be composed of individual chemical reactions. Each reaction is itself an input-output subsystem.  The input signal arises from the extrinsic spatial and temporal fluctuations of chemical concentrations.  The output follows from the chemical transformations of the reaction that alter concentrations.  The overall input-output pattern of the system develops from the signal processing of the component subsystems and the aggregate architecture of the components that form the broader system. 

Many fundamental questions in biology come down to understanding these input-output relations. Some systems are broadly sensitive, changing outputs moderately over a wide range of inputs.  Other systems are ultrasensitive or bistable, changing very rapidly from low to high output across a narrow range of inputs \autocite{tyson03sniffers}.  The Hill equation describes these commonly observed input-output patterns, capturing the essence of how changing inputs alter system response \autocite{zhang13ultrasensitive}.  

I start with two key questions.  How does the commonly observed ultrasensitive response emerge, given that classical chemical kinetics does not naturally lead to that pattern? Why does the very simple Hill equation match so well to the range of observed input-output relations?  To answer those questions, I emphasize the general processes that shape input-output relations.  Three aspects seem particularly important: aggregation, measurement, and scale.

Aggregation combines lower-level processes to produce the overall input-output pattern of a system.  Aggregation often transforms numerous distinct and sometimes disordered lower-level fluctuations into highly regular overall pattern \autocite{frank09the-common}.  One must understand those regularities in order to analyze the relations between pattern and process. Aggregate regularity also imposes constraints on how natural selection shapes biological design \autocite{kauffman93the-origins}.

Measurement describes the information captured from inputs and transmitted through outputs.  How sensitive are outputs to a change in inputs?  The overall pattern of sensitivity affects the information lost during measurement and the information that remains invariant between input and output. Patterns of sensitivity that may seem puzzling or may appear to be specific to particular mechanisms often become simple to understand when one learns to read the invariant aspects of information and measurement. Measurement also provides a basis for understanding scale \autocite{hand04measurement}.

Scale influences the relations between input and output \autocite{stevens57on-the-psychophysical}.  Large input typically saturates a system, causing output to become insensitive to further increases in input.  Saturated decline in sensitivity often leads to logarithmic scaling.  Small input often saturates in the other direction, such that output changes slowly and often logarithmically in response to further declines in input.  The Hill equation description of input-output patterns is simply an expression of logarithmic saturation at high and low inputs, with an increased linear sensitivity at intermediate input levels.

High input saturates output because maximum output is intrinsically limited.  By contrast, the commonly observed logarithmic saturation at low input intensity remains a puzzle.  The difficulty arises because typical theoretical understanding of chemical kinetics predicts a strong and nearly linear output sensitivity at low input concentrations of a signal \autocite{cornish-bowden12fundamentals}. That theoretical linear sensitivity of chemical kinetics at low input contradicts the widely observed pattern of weak logarithmic sensitivity at low input.  

I describe the puzzle of chemical kinetics in the next section to set the basis for a broader analysis of input-output relations. I then connect the input-output relations of chemical kinetics to universal aspects of aggregation, measurement, and scale. Those universal properties of input-output systems combine with specific biological mechanisms to determine how biological systems respond to inputs. Along the way, I consider possible resolutions to the puzzle of chemical kinetics and to a variety of other widely observed but unexplained regularities in input-output patterns.  Finally, I discuss the ways in which regularities of input-output relations shape many key aspects of biological design.

\section*{Review}

\subsection*{The puzzle of chemical kinetics}

Classical Michaelis-Menten kinetics for chemical reactions lead to a saturating relationship between an input signal and an output response \autocite{cornish-bowden12fundamentals}.  The particular puzzle arises at very low input, for which Michaelis-Menten kinetics predict a nearly linear output response to tiny changes in input.  That sensitivity at low input means that chemical reactions would have nearly infinite measurement precision with respect to tiny fluctuations of input concentration. Idealized chemical reactions do have that infinite precision, and observations may follow that pattern if nearly ideal conditions are established in laboratory studies.  By contrast, the actual input-output relations of chemical reactions and more complex biological signals often depart from Michaelis-Menten kinetics.

Many studies have analyzed the contrast between Michaelis-Menten kinetics and the observed input-output relations of chemical reactions \autocite{zhang13ultrasensitive}.  I will discuss some of the prior studies in a later section.  However, before considering those prior studies, it is useful to have a clearer sense of the initial puzzle and of alternative ways in which to frame the problem.

\subsubsection*{Example of Michaelis-Menten kinetics}

I illustrate Michaelis-Menten input-output relations with a particular example, in which higher input concentration of a signal increases the transformation of an inactive molecule to an active state. Various formulations of Michaelis-Menten kinetics emphasize different aspects of reactions \autocite{cornish-bowden12fundamentals}. But those different formulations all have the same essential mass action property that assumes spatially independent concentrations of reactants.  Spatially independent concentrations can be multiplied to calculate the spatial proximity between reactants at any point in time. 

In my example, a signal, S, changes an inactive reactant, R, to an active output, A, in the reaction
\begin{equation*}
  \ce{S + R ->[g] S + A},
\end{equation*}
where the rate of reaction, $g$, can be thought of as the signal gain.  In this reaction alone, if $S>0$, all of the reactant, $R$, will eventually be transformed into the active form, $A$. (I use roman typeface for the distinct reactant species and italic typeface for concentrations of those reactants.)  However, I am particularly interested in the relation between the input signal concentration, $S$, and the output signal concentration, $A$.  Thus, I also include a back reaction, in which the active form, A, spontaneously transforms back to the inactive form, R, expressed as
\begin{equation*}
  \cee{A ->[\Gd] R}.
\end{equation*}
The reaction kinetics follow
\begin{equation}\label{eq:mmKinetics}
  \dot{A} = gS(N-A) - \Gd A,
\end{equation}
in which the overdot denotes the derivative with respect to time, and $N=R+A$ is the total concentration of inactive plus active reactant molecules. We find the equilibrium concentration of the output signal, $A^*$, as a function of the input signal, $S$, by solving $\dot{A}=0$, which yields
\begin{equation}\label{eq:mmEq}
  A^* = N\left(\frac{S}{m+S}\right),
\end{equation}
in which $m=\Gd/g$ is the rate of the back reaction relative to the forward reaction.  Note that $S/(m+S)$ is the equilibrium fraction of the initially inactive reactant that is transformed into the active state.  At $S=m$, the input signal transforms one-half of the inactive reactant into the active state.  

\begin{figure}[t!]
\centering
\includegraphics[width=\hsize]{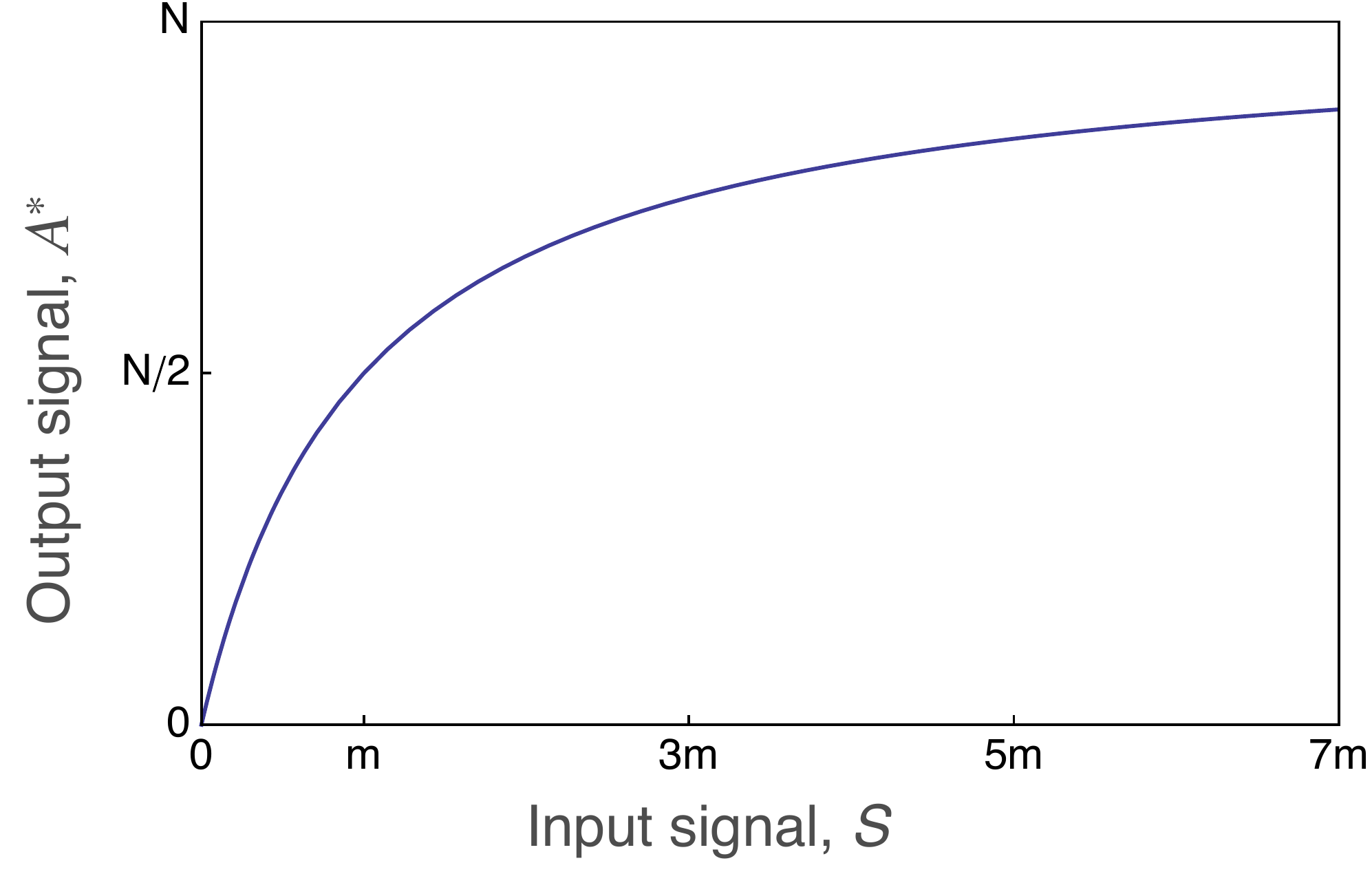}
\caption{\csentence{Michaelis-Menten signal transmission.} The reaction dynamics transform the concentration of the input signal, $S$, into the equilibrium output signal, $A^*$, as given by \Eq{mmEq}. Half maximal output occurs at input $S=m$.  The total reactant available to be transformed is $N$.
\hbox{\null}
\label{fig:mmExample}
}
\end{figure}

\Fig{mmExample} shows the consequence of this type of Michaelis-Menten kinetics for the relation between the input signal and the output signal.  At low input signal intensity, $S\rightarrow0$, the output is strongly (linearly) sensitive to changes in input, with the output changing in proportion to $S$. At high signal intensity, the output is weakly (logarithmically) sensitive to changes in input, with the output changing in proportion to $\log(S)$.  The output saturates at $A^*\rightarrow N$ as the input increases.

\subsubsection*{The Hill equation and observed input-output patterns}

Observed input-output patterns often differ from the simple Michaelis-Menten pattern in \Fig{mmExample}.  In particular, output is often only weakly sensitive to changes in the input signal at low input intensity. Weak sensitivity at low input values often means that output changes in proportion to $\log(S)$ for small $S$ values, rather than the linear relation between input and output at small $S$ values described by Michaelis-Menten kinetics.

\begin{table}[t!]
\caption{Conceptual foundations.}\label{box:psycho}
\begin{flushleft}
\parindent=12pt
Psychophysics studies human perception of quantity, such as loudness, temperature or pressure.  The early work of Weber and Fechner suggested that perception scales logarithmically: for a given stimulus (input), the perception of quantity (output) changes logarithmically.  That work led to modern analysis of measurement and scale.  

This article analyzes biological input-output relations.  My examples focus on biochemistry.  I chose that focus for two reasons.  First, most biological input-output relations may ultimately be reducible to cascades of biochemical component reactions.  The problem then becomes how to relate the biochemical components and their connections to overall system function.  That relation between biochemistry and system function is the core of modern systems biology.  Second, the sharp distinction between classical Michaelis-Menten chemical kinetics and the observed patterns of logarithmic scaling in both biochemistry and perception provides a good entry into the unsolved puzzles of the subject and the potential value of my perspective.

Although I focus on biochemistry, my approach derives from other topics.  I borrow the deep conceptual foundations of measurement from psychophysics, the principles of aggregation from statistical mechanics, and aspects of information theory that originally developed in studies of communication.  My view is that biological input-output relations can only be understood in terms of aggregation, measurement and information.  In this article, I evoke those principles indirectly by building a series of specific analyses of biochemistry and simple aspects of systems biology.

The literatures and conceptual spans are vast for psychophysics, measurement theory, statistical mechanics and information theory.  Here, I mention a few key entries into each subject.  To read this article, it is not necessary to understand all of those topics.  But it is necessary to see the project for what it is, an attempt to borrow deep principles from other subjects and apply those principles to biochemical aspects of systems biology, to the nature of biological input-output relations, and to the consequences for natural selection and evolutionary design.

Gescheider \autocite{gescheider97psychophysics:} summarizes aspects of psychophysics related to my discussion of input-output patterns. History and further references can be obtained from that work.  Certain aspects of measurement theory followed from psychophysics \autocite{stevens57on-the-psychophysical,hand04measurement}. The theory developed into a broader analysis of the principles of quantity \autocite{krantz06foundations,krantz06foundationsb,suppes06foundations}.  Other branches of measurement theory focus on aspects of precision and calibration \autocite{rabinovich05measurement}.

Statistical mechanics analyzes the ways in which aggregation leads to highly ordered systems arising from disordered underlying components.  My usage follows from the proposed unity between information theory and aggregate pattern, which transcends the specifics of physical models and instead emphasizes the patterns expressed by probability distributions \autocite{jaynes03probability,frank09the-common}. That Jaynesian perspective describes how aggregation dissipates information to expose underlying regularity. Later work \autocite{frank10measurement,frank11a-simple} provided a unified framework for all common probability patterns by combining measurement theory with Jaynes' information theory interpretation of statistical mechanics.
\end{flushleft}
\end{table}

The Hill equation preserves the overall Michaelis-Menten pattern but alters the sensitivity at low inputs to be logarithmic rather than linear. Remarkably, the pattern of curve shapes for most biochemical reactions and more general biological input-output relations fit reasonably well to the Hill equation
\begin{equation}\label{eq:hilldim}
  \hat{y} = b\left(\frac{\hat{x}^k}{m^k+\hat{x}^k}\right)
\end{equation}
or to minor variants of this equation (\Table{psycho}). The input intensity is $\hat{x}$, the measured output is $\hat{y}$, half of maximal response is $\hat{x}=m$, the shape of the response is determined by the Hill coefficient $k$, and the response saturates asymptotically at $b$ for increasing levels of input.  

\begin{figure}[t!]
\centering
\includegraphics[width=\hsize]{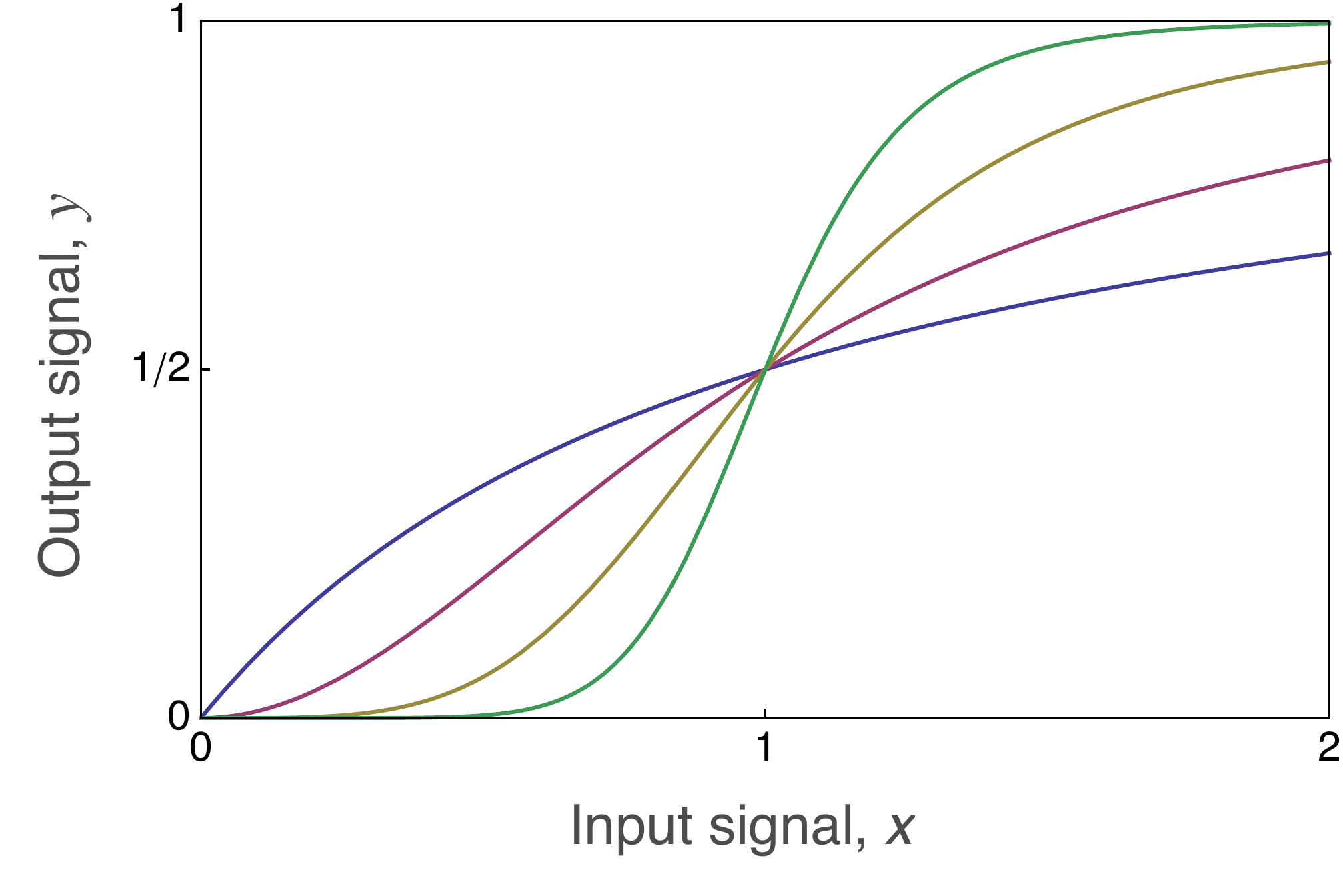}
\caption{\csentence{Hill equation signal transmission.} The input signal, $x$, leads to the output, $y$, as given by \Eq{hill}. The curves of increasing slope correspond to $k=1,2,4,8$.  
\hbox{\null}
\label{fig:hillEx}
}
\end{figure}

We can simplify the expression by using the substitutions $y=\hat{y}/b$, in which $y$ is the fraction of the maximal response, and $x=\hat{x}/m$, in which $x$ is the ratio of the input to the value that gives half of the maximal response.  The resulting equivalent expression is 
\begin{equation}\label{eq:hill}
  y = \frac{x^k}{1+x^k}.
\end{equation}
\Fig{hillEx} shows the input-output relations for different values of the Hill coefficient, $k$. For $k=1$, the curve matches the Michaelis-Menten pattern in \Fig{mmExample}.  An increase in $k$ narrows the input range over which the output responds rapidly (sensitively).  For larger values of $k$, the rapid switch from low to high output response is often called a bistable response, because the output state of the system switches in a nearly binary way between low output, or ``OFF'', and high output, or ``ON''. A bistable switching response is effectively a biological transistor that forms a component of a biological circuit \autocite{sarpeshkar10ultra}. Bistability is sometimes called ultrasensitivity, because of the high sensitivity of the response to small changes in inputs when measured over the responsive range \autocite{goldbeter81an-amplified}.

\begin{figure}[t!]
\centering
\includegraphics[width=\hsize]{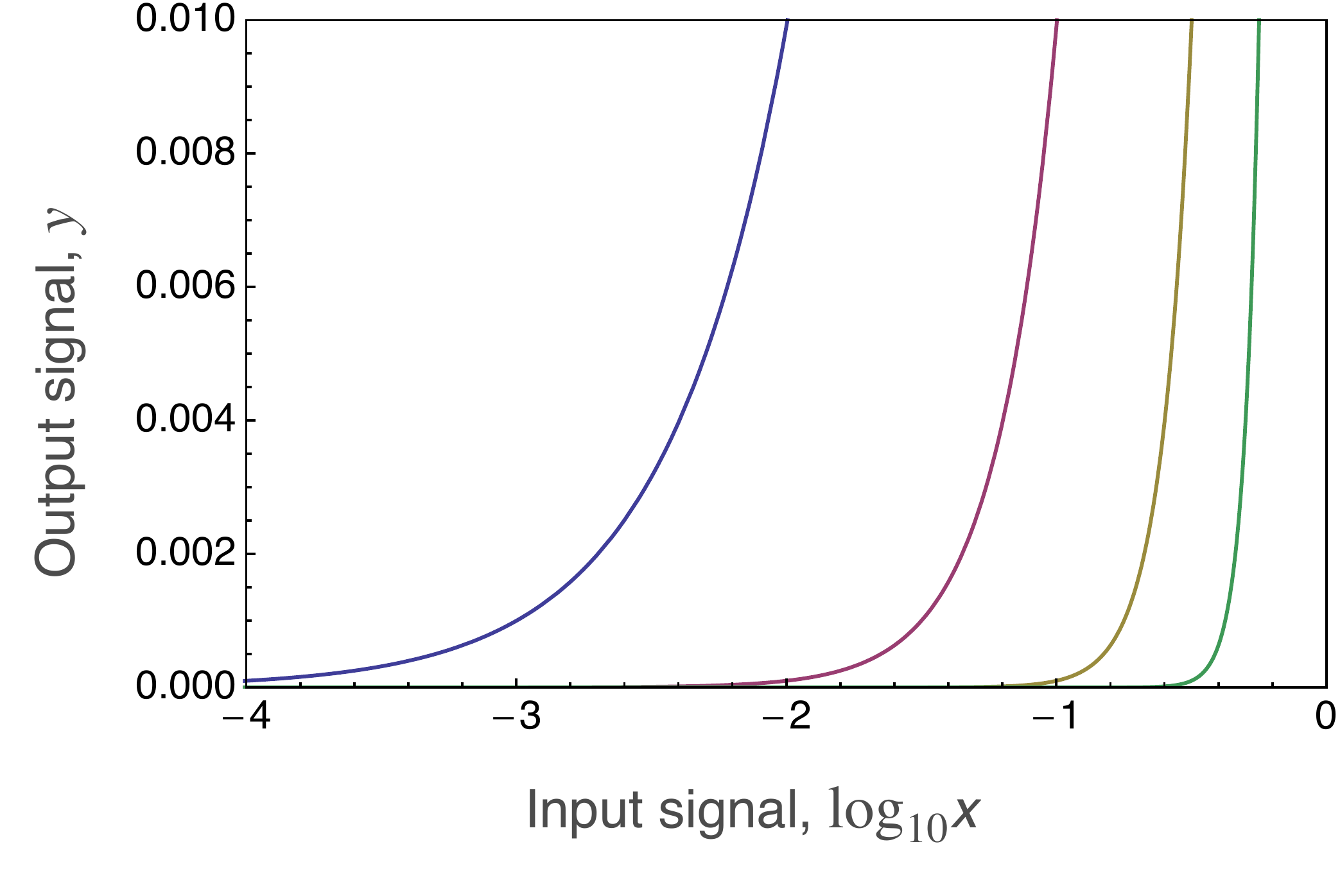}
\caption{\csentence{An increasing Hill coefficient, $k$, causes logarithmic sensitivity to low input signals.}  At $k=1$ (left curve), the sensitivity is linear with a steady increase in output even at very low input levels, implying infinite precision.  As $k$ increases, sensitivity at low input declines, and the required threshold input level becomes higher and sharper to induce an output response of $1\%$ of the maximum $(y=0.01)$. The curves of increasing slope correspond to $k=1,2,4,8$ in \Eq{hill}, with logarithmic scaling of the input $x$ plotted here.  
\hbox{\null}
\label{fig:logIn}
}
\end{figure}

At the $k=1$ case of Michaelis-Menten, the output response is linearly sensitive to very small changes at very low input signals.  Such extreme sensitivity means essentially infinite measurement precision at tiny input levels, which seems unlikely for realistic biological systems.  As $k$ increases, sensitivity at low input becomes more like a threshold response, such that a minimal input is needed to stimulate significant change in output.  Increasing $k$ causes sensitivity to become logarithmic at low input. That low input sensitivity pattern can be seen more clearly by plotting the input level on a logarithmic scale, as in \Fig{logIn}.

\subsection*{Alternative mechanisms for simple chemical reactions}

My goal is to understand the general properties of input-output relations in biological systems.  To develop that general understanding, it is useful to continue with study of the fundamental input-output relations of simple chemical reactions.  Presumably, most input-output relations of systems can ultimately be decomposed into simple component chemical reactions.  Later, I will consider how the combination of such components influences overall system response.

Numerous studies of chemical kinetics report Hill coefficients $k>1$ rather than the expected Michaelis-Menten pattern $k=1$.  Resolution of that puzzling discrepancy is the first step toward deeper understanding of input-output patterns (\Table{hill}). Zhang et al.\ \autocite{zhang13ultrasensitive} review six specific mechanisms that may cause $k>1$.  In this section, I briefly summarize several of those mechanisms. See Zhang et al.\ \autocite{zhang13ultrasensitive} for references.

\begin{table}[t!]
\caption{Literature related to the Hill equation.}\label{box:hill}
\begin{flushleft}\parindent=12pt
The Hill equation or related expressions form a key part of analysis in many areas of chemistry, systems biology and pharmacology.  Each subdiscipline has its own set of isolated conceptual perspectives and mutual citation islands. Here, I list a few publications scattered over that landscape.  My haphazard sample provides a sense of the disconnected nature of the topic.  I conclude from this sample that the general form of input-output relations is widely recognized as an important problem and that no unified conceptual approach exists.  

Cornish-Bowden's \citeyear{cornish-bowden12fundamentals} text on enzyme kinetics frequently uses the Hill coefficient to summarize the relation between input concentrations and the rate of transformation to outputs. That text applies both of the two main approaches. First, the Hill equation simply provides a description of how changes in input affect output independently of the underlying mechanism. Second, numerous specific models attempt to relate particular mechanisms to observed Hill coefficients.  Zhang et al.\ \autocite{zhang13ultrasensitive} provide an excellent, concise summary of specific biochemical mechanisms, including some suggested connections to complex cellular phenotypes.  

Examples of the systems biology perspective include Kim \& Ferrell \autocite{kim07substrate}, Ferrell \autocite{ferrell09signaling}, Cohen-Saidon et al.\ \autocite{cohen-saidon09dynamics}, Goentoro \& Kirschner \autocite{goentoro09evidence} and Goentoro et al.\ \autocite{goentoro09the-incoherent}.  Alon's \citeyear{alon07an-introduction} leading text in systems biology discusses the importance of the Hill equation pattern, but only considers the explicit classical chemical mechanism of multiple binding.  Those studies share the view that specific input-output patterns require specific underlying mechanisms as explanations. 

In pharmacology, the Hill equation provides the main approach for describing dose-response patterns.  Often, the Hill equation is used as a model to fit the data independently of mechanism. That descriptive approach probably follows from the fact that many complex and unknown factors influence the relation between dose and response.  Alternatively, some analyses focus on the key aspect of receptor-ligand binding in the response to particular drugs. Reviews from this area include DeLean et al.\ \autocite{delean78simultaneous}, Weiss \autocite{weiss97the-hill}, Rang \autocite{rang06the-receptor} and Bindslev \autocite{bindslev08drug-acceptor}.  Related approaches arise in the analysis of basic physiology \autocite{walker10analysing}.

Other approaches consider input-output responses in relation to aggregation of underlying heterogeneity, statistical mechanics or aspects of information.  Examples include Hoffman \& Goldberg \autocite{hoffman94the-relationship}, Getz \& Lansky \autocite{getz01receptor}, Kolch et al.\ \autocite{kolch05when}, Tka{\v{c}}ik \& Walczak \autocite{tkacik11information} and Marzen et al.\ \autocite{marzen13statistical}. 

Departures from the mass-action assumption of Michaelis-Menten kinetics can explain the emergence of Hill equation input-output relations \autocite{savageau95michaelis-menten,savageau98development}.  Many studies analyze the kinetics of diffusion-limited departures from mass action without making an explicit connection to the Hill equation \autocite{ben-avraham00diffusion,andrews04stochastic,schnell04reaction}. Modeling approaches in other disciplines also consider the same problem of departures from spatial uniformity \autocite{dieckmann00the-geometry,ellner01pair,marro05nonequilibrium}

Studies often use the Hill equation or similar assumptions to describe the shapes of input-output functions when building models of biochemical circuits \autocite{kholodenko97quantification,sarpeshkar10ultra}.  Those studies do not make any mechanistic assumptions about the underlying cause of the Hill equation pattern.  Rather, in order to build a model circuit for regulatory control, one needs to make some assumption about input-output relations. 
\end{flushleft}
\end{table}

\subsubsection*{Direct multiplication of signal input concentration}

Transforming a single molecule to an active state may require simultaneous binding by multiple input signal molecules.  If two signal molecules, S, must bind to a single inactive reactant, R, to form a three molecule complex before transformation of R to the active state, A, then we can express the reaction as 
\begin{equation*}
  \ce{S + S + R ->[g] SSR -> S + S + A},
\end{equation*}
which by mass action kinetics leads to the rate of change in A as
\begin{equation*}
  \dot{A} = gS^2(N-A) - \Gd A,
\end{equation*}
in which $N=R+A$ is the total concentration of the inactive plus active reactant molecules, and the back reaction $A\rightarrow R$ occurs at rate $\Gd$. The equilibrium input-output relation is
\begin{equation*}
  A^* = N\left(\frac{S^2}{m^2+S^2}\right),
\end{equation*}
which is a Hill equation with $k=2$.  The reaction stoichiometry, with two signal molecules combining in the reaction, causes the reaction rate to depend multiplicatively on signal input concentration. Other simple schemes also lead to a multiplicative effect of signal molecule concentration on the rate of reaction. For example, the signal may increase the rates of two sequential steps in a pathway, causing a multiplication of the signal concentration in the overall rate through the multiple steps.  Certain types of positive feedback can also amplify the input signal multiplicatively.

\subsubsection*{Saturation and loss of information in multistep reaction cascades}

The previous section discussed mechanisms that multiply the signal input concentration to increase the Hill coefficient.  Multiplicative interactions lead to logarithmic scaling.  The Hill equation with $k>1$ expresses logarithmic scaling of output at high and low input levels.  I will return to this general issue of logarithmic scaling later.  The point here is that multiplication is one sufficient way to achieve logarithmic scaling.  But multiplication is not necessary.  Other nonmultiplicative mechanisms that lead to logarithmic scaling can also match closely to the Hill equation pattern. This section discusses two examples covered by Zhang et al.\ \autocite{zhang13ultrasensitive}.

\paragraph*{Repressor of weak input signals}

The key puzzle of the Hill equation concerns how to generate the logarithmic scaling pattern at low input intensity.  The simplest nonmultiplicative mechanism arises from an initial reaction that inactivates the input signal molecule. That preprocessing of the signal intensity can create a filter that logarithmically reduces signals of low intensity.  Suppose, for example, that the repressor may become saturated at higher input concentrations. Then the initial reaction filters out weak, low concentration, inputs but passes through higher input concentrations.  

\begin{figure}[t!]
\centering
\includegraphics[width=\hsize]{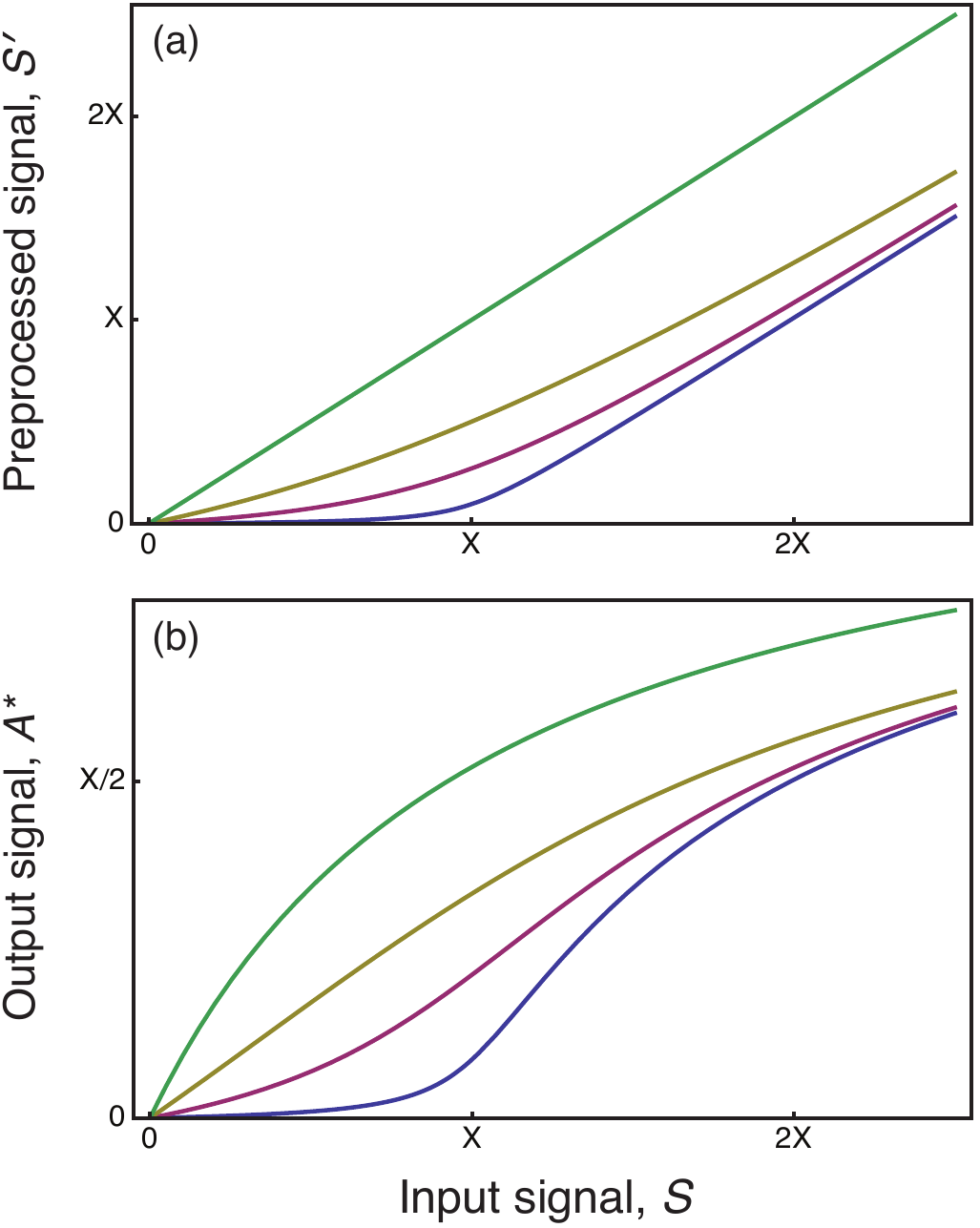}
\caption{\csentence{Preprocessing of an input signal by a repressor reduces sensitivity of output to low input intensity signals.}   (a) Equilibrium concentration of processed signal, $S^\prime$, in relation to original signal input intensity, $S$, obtained by solution of \Eq{preKinetics}.   The four curves from bottom to top show decreasing levels of signal titration by the repressor for the parameter values $\Ga=0.01,0.1,0.5,1000$.  The top curve alters the initial signal very little, so that $S^\prime\approx S$, showing the consequences of an unfiltered input signal.  (b) The processed input signal, $S^\prime$, is used as the input to a standard Michaelis-Menten reaction kinetics process in \Eq{mmKinetics}, leading to an equilibrium output, $A^*$.  The curves from bottom to top derive from the corresponding preprocessed input signal from the upper panel. 
\hbox{\null}
\label{fig:preprocess}
}
\end{figure}

Consider a repressor, X, that can bind to the signal, S, transforming the bound complex into an inactive state, I, in the reaction
\begin{equation*}
  \cee{S + X <=>[\Gg][\Gb] I}.
\end{equation*}
One can think of this reaction as a preprocessing filter for the input signal.  The kinetics of this input preprocessor can be expressed by focusing on the change in the concentration of the bound, inactive complex
\begin{equation}\label{eq:preKinetics}
  \dot{I} = \Gg(S-I)(X-I) - \Gb I.
\end{equation}
The signal passed through this preprocessor is the amount of S that is not bound in I complexes, which is $S'=S-I$.  We can equivalently write $I=S'-S$.  The equilibrium relation between the input, $S$, and the output signal, $S'$, passed through the preprocessor can be obtained by solving $\dot{I}=0$, which yields
\begin{equation*}
  S'(X-S+S') - \Ga(S-S')=0,
\end{equation*}
in which $\Ga=\Gb/\Gg$.  \Fig{preprocess}a shows the relation between the input signal, $S$, and the preprocessed output, $S'$.  Bound inactive complexes, $I$, hold the signal molecule tightly and titrate it out of activity when the breaking up of complexes at rate $\Gb$ is slower than the formation of new complexes at rate $\Gg$, and thus $\Ga$ is small.  

The preprocessed signal may be fed into a standard Michaelis-Menten type of reaction, such as the reaction in \Eq{mmKinetics}, with the preprocessed signal $S^\prime$ driving the kinetics rather than the initial input, $S$.  The reaction chain from initial input through final output starts with an input concentration, $S$, of which $S^\prime$ passes through the repressor filter, and $S^\prime$ stimulates production of the active output signal concentration, $A^*$. \Fig{preprocess}b shows that titration of the initial signal concentration, $S$, to a lower pass-through signal concentration, $S^\prime$, leads to low sensitivity of the final output, $A^*$, to the initial signal input, $S$, as long as the signal concentration is below the amount of the repressor available for titration, $X$.  

When this signal preprocessing mechanism occurs, the low, essentially logarithmic, sensitivity to weak input signals solves the puzzle of relating classical Michaelis-Menten chemical kinetics to the Hill equation pattern for input-output relations with $k>1$.  The curves in \Fig{preprocess}b do not exactly match the Hill equation. However, this signal preprocessing mechanism aggregated with other simple mechanisms can lead to a closer fit to the Hill equation pattern.  I discuss the aggregation of different mechanisms below. 

This preprocessed signal system is associated with classical chemical kinetic mechanisms, because it is the deterministic outcome of a simple and explicit mass action reaction chain.  However, the reactions are not inherently multiplicative with regard to signal input intensity. Instead, preprocessing leads to an essentially logarithmic transformation of scaling and information at low input signal intensity. 

This example shows that the original notion of multiplicative interactions is not a necessary condition for Hill equation scaling of input-output relations.  Instead, the Hill equation pattern is simply a particular expression of logarithmic scaling of the input-output relation.  Any combination of processes that leads to similar logarithmic scaling provides similar input-output relations.  Thus, the Hill equation pattern does not imply any particular underlying chemical mechanism.  Rather, such input-output relations are the natural consequence of the ways in which information degrades and is transformed in relation to scale when passed through reaction sequences that act as filters of the input signal.

\paragraph*{Opposing forward and back reactions}

The previous section showed how a repressor can reduce sensitivity to low intensity input signals.  A similar mechanism occurs when there is a back reaction.  For example, a signal may transform an inactive reactant into an active form, and a back reaction may return the active form to the inactive state.  If the back reaction saturates at low signal input intensity, then a rise in the signal from a very low level will initially cause relatively little increase in the concentration of the active output, inducing weak, logarithmic sensitivity to low input signal intensity.  In effect, the low input is repressed, or titrated, by the strong back reaction.

This opposition between forward and back reactions was one of the first specific mechanisms of classical chemical kinetics to produce the Hill equation pattern in the absence direct multiplicative interactions that amplify the input signal \autocite{goldbeter81an-amplified}.  In this section, I briefly illustrate the opposition of forward and back reactions in relation to the Hill equation pattern.

In the forward reaction, a signal, S, transforms an inactive reactant, R, into an active state, A.  The back reaction is catalyzed by the molecule B, which transforms A back into R.  The balancing effects of the forward and back reactions in relation to saturation depend on a more explicit expression of classical Michaelis-Menten kinetics than presented above.  In particular, let the two reactions be
\begin{align*}
  \cee{S + R &<=>[g][\Gd] SR ->[\Gf] S + A} \\
  \cee{B + A &<=>[\Gg][d] BA ->[\Gs] B + R}
\end{align*}
in which these reactions show explicitly the intermediate bound complexes, SR and BA. The rate of change in the output signal, $\dot{A}$, when the dynamics follow classical equilibrium Michaelis-Menten reaction kinetics, is
\begin{equation}\label{eq:forwardback}
  \dot{A} = \Gf S_0\left(\frac{R}{m+R}\right)- \Gs B_0\left(\frac{A}{\Gm+A}\right),
\end{equation}
in which $S_0$ includes the concentrations of both free signal, S, and bound signal, SR. Similarly, $B_0$ includes the concentrations of both free catalyst, B, and bound catalyst, BA.  The half-maximal reaction rates are set by $m=\Gd/g$ and $\Gm=d/\Gg$. The degree of saturation depends on the total amount of reactant available, $N=R+A$, relative to the concentrations that give the half-maximal reaction rates, $m$ and $\Gm$.    

\begin{figure}[t!]
\centering
\includegraphics[width=\hsize]{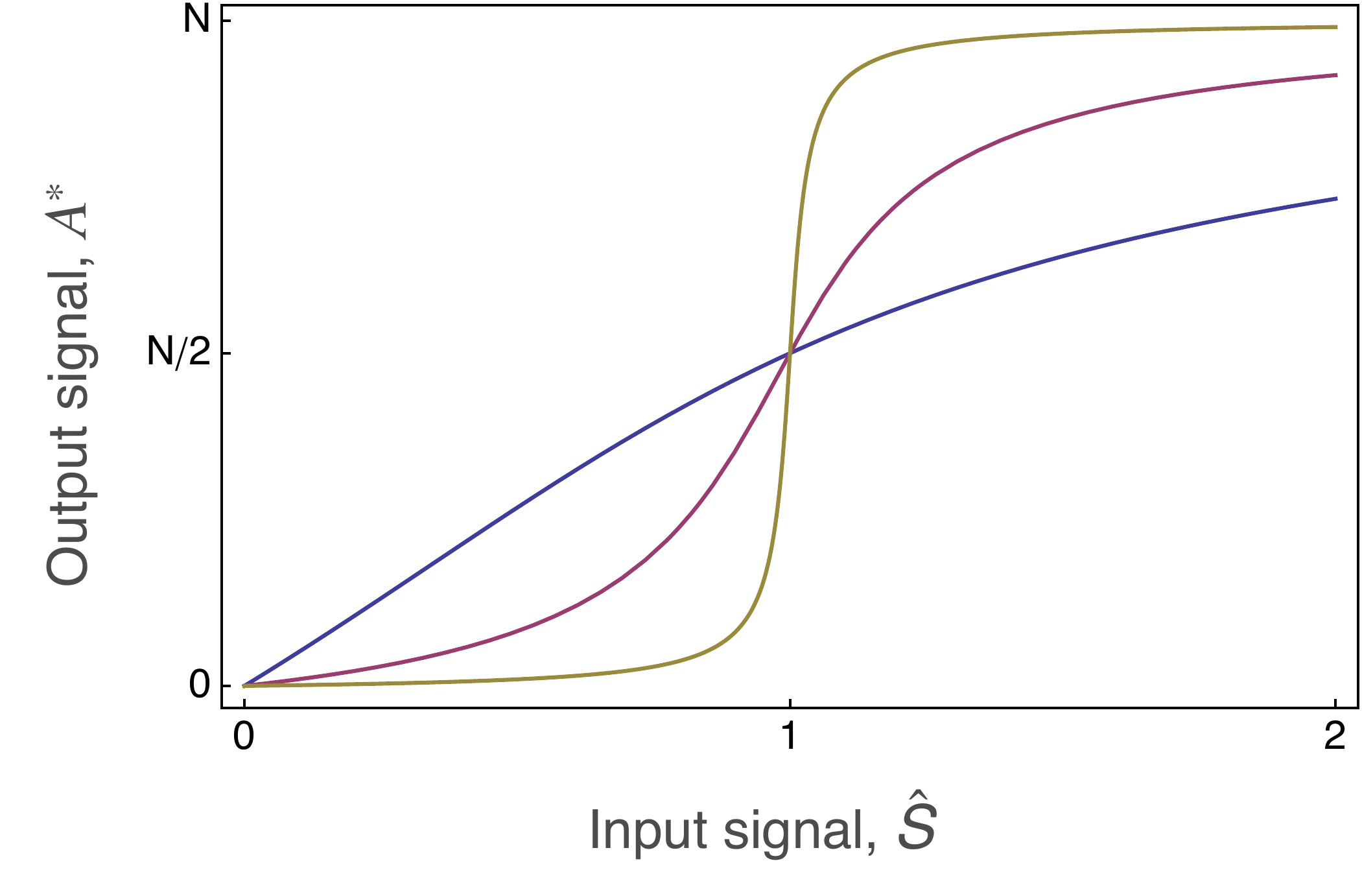}
\caption{\csentence{Balance between forward and back reactions leads to a high Hill coefficient when the reactions are saturated.}  The equilibrium output signal, $A^*$, is obtained by solving $\dot{A}=0$ in \Eq{forwardback} as a function of the input signal, $S_0$.  The signal is given as $\hat{S}=\Gf S_0/\Gs B_0$.  The total amount of reactant is $N=R+A$. The half-maximal concentrations are set to $m=\Gm=1$.  The three curves illustrate the solutions for $N=1,10,100$, with increasing Hill coefficients for higher $N$ values and greater reaction saturation levels.
\hbox{\null}
\label{fig:forwardback}
}
\end{figure}

When the input signal, $S_0$, is small, the back reaction dominates, potentially saturating the forward rate as $R$ becomes large.  \Fig{forwardback} shows that the level of saturation sets the input-output pattern, with greater saturation increasing the Hill coefficient, $k$. 

\subsection*{Alternative perspectives on input-output relations}

In the following sections, I discuss alternative mechanisms that generate Hill equation patterns.  Before discussing those alternative mechanisms, it is helpful to summarize the broader context of how biochemical and cellular input-output relations have been studied.

\subsubsection*{Explicit chemical reaction mechanisms} 

The prior sections linked simple and explicit chemical mechanisms to particular Hill equation patterns of inputs and outputs.  Each mechanism provided a distinct way in which to increase the Hill coefficient above one.  Many key reviews and textbooks in biochemistry and systems biology emphasize that higher Hill coefficients and increased input-output sensitivity arise from these simple and explicit deterministic mechanisms of chemical reactions \autocite{alon07an-introduction,cornish-bowden12fundamentals,zhang13ultrasensitive}. The idea is that a specific pattern must be generated by one of a few well-defined and explicit alternative mechanisms.  

Explicit chemical reaction mechanisms discussed earlier include: binding of multiple signal molecules to stimulate each reaction; repressors of weak input signals; and opposing forward and back reactions near saturation.  Each of these mechanisms could, in principle, be isolated from a particular system, analyzed directly, and linked quantitatively to the specific input-output pattern of a system.  Decomposition to elemental chemical kinetics and direct quantitative analysis would link observed pattern to an explicit mechanistic process.

\subsubsection*{The Hill equation solely as a description of observed pattern} 

In the literature, the Hill equation is also used when building models of how system outputs may react to various inputs (\Table{hill}). The models often study how combinations of components lead to the overall input-output pattern of a system.  To analyze such models, one must make assumptions about the input-output relations of the individual components.  Typically, a Hill equation is used to describe the components' input-output functions.  That description does not carry any mechanistic implication.  One simply needs an input-output function to build the model or to describe the component properties.  The Hill equation is invoked because, for whatever reason, most observed input-output functions follow that pattern.

\subsubsection*{System-level mechanisms and departures from mass action} 

Another line of study focuses on system properties rather than the input-output patterns of individual components.  In those studies, the Hill equation pattern of sensitivity does not arise from a particular chemical mechanism in a particular reaction.  Instead, sensitivity primarily arises from the aggregate consequences of the system (\Table{hill}).  In one example, numerous reactions in a cascade combine to generate Hill-like sensitivity  \autocite{kholodenko97quantification}. The sensitivity derives primarily from the haphazard combination of different scalings in the distinct reactions, rather than a particular chemical process.  

Alternatively, some studies assume that chemical kinetics depart from the classical mass action assumption (\Table{hill}). If input signal molecules tend, over the course of a reaction, to become spatially isolated from the reactant molecules on which they act, then such spatial processes often create a Hill-like input-output pattern by nonlinearly altering the output sensitivity to changes in inputs.  I consider such spatial processes as an aggregate system property rather than a specific chemical mechanism, because many different spatial mechanisms can restrict the aggregate movement of molecules.  The aggregate spatial processes of the overall system determine the departures from mass action and the potential Hill-like sensitivity consequences, rather than the particular physical mechanisms that alter spatial interactions.  

These system-level explanations based on reaction cascades and spatially induced departures from mass action have the potential benefit of applying widely.  Yet each particular system-level explanation is itself a particular mechanism, although at a higher level than the earlier biochemical mechanisms. In any actual case, the higher system-level mechanism may or may not apply, just as each explicit chemical mechanism will sometimes apply to a particular case and sometimes not.  

\subsubsection*{A broader perspective} 

As we accumulate more and more alternative mechanisms that fit the basic input-output pattern, we may ask whether we are converging on a full explanation or missing something deeper. Is there a different way to view the problem that would unite the disparate perspectives, without losing the real insights provided in each case?

I think there is a different, more general perspective (\Table{psycho}).  At this point, I have given just enough background to sketch that broader perspective.  I do so in the remainder of this section.  However, it is too soon to go all the way.  After giving a hint here about the final view, I return in the following sections to develop further topics, after which I return to a full analysis of the broader ways in which to think about input-output relations.

The Hill equation with $k>1$ describes weak, logarithmic sensitivity at low input and high input levels, with strong and essentially linear sensitivity through an intermediate range.  Why should this log-linear-log pattern be so common?  The broader perspective on this problem arises from the following points.

First, the common patterns of nature are exactly those patterns consistent with the greatest number of alternative underlying processes \autocite{jaynes03probability,frank09the-common}.  If many different processes lead to the same outcome, then that outcome will be common and will lack a strong connection to any particular mechanism.  In any explicit case, there may be a simple and clear mechanism.  But the next case, with the same pattern, is likely to be mechanistically distinct.

Second, measurement and information transmission unite the disparate mechanisms.  The Hill equation with $k>1$ describes a log-linear-log measurement scale \autocite{frank10measurement,frank11a-simple}.  The questions become: Why do biological systems, even at the lowest chemical levels of analysis, often follow this measurement scaling?  How does chemistry translate into the transmission and loss of information in relation to scale?  Why does a universal pattern of information and measurement scale arise across such a wide range of underlying mechanisms?

Third, this broader perspective alters the ways in which one should analyze biological input-output systems.  In any particular case, specific mechanisms remain interesting and important. However, the relations between different cases and the overall interpretation of pattern must be understood within the broader framing. 

With regard to biological design, natural selection works on the available patterns of variation.  Because certain input-output relations tend to arise, natural selection works on variations around those natural contours of input-output patterns. Those natural contours of pattern and variation are set by the universal properties of information transmission and measurement scale. That constraint on variation likely influences the kinds of designs created by natural selection.  To understand why certain designs arise and others do not, we must understand how information transmission and measurement scale set the underlying patterns of variation.

I return to these points later.  For now, it is useful to keep in mind these preliminary suggestions about how the various pieces will eventually come together.

\subsection*{Aggregation}

Most biological input-output relations arise through a series of reactions.  Initial reactions transform the input signal into various intermediate signals, which themselves form inputs for further reactions. The final output arises only after multiple internal transformations of the initial signal.  We may think of the overall input-output relation as the aggregate consequence of multiple reaction components.  

\begin{figure}[t!]
\centering
\includegraphics[width=\hsize]{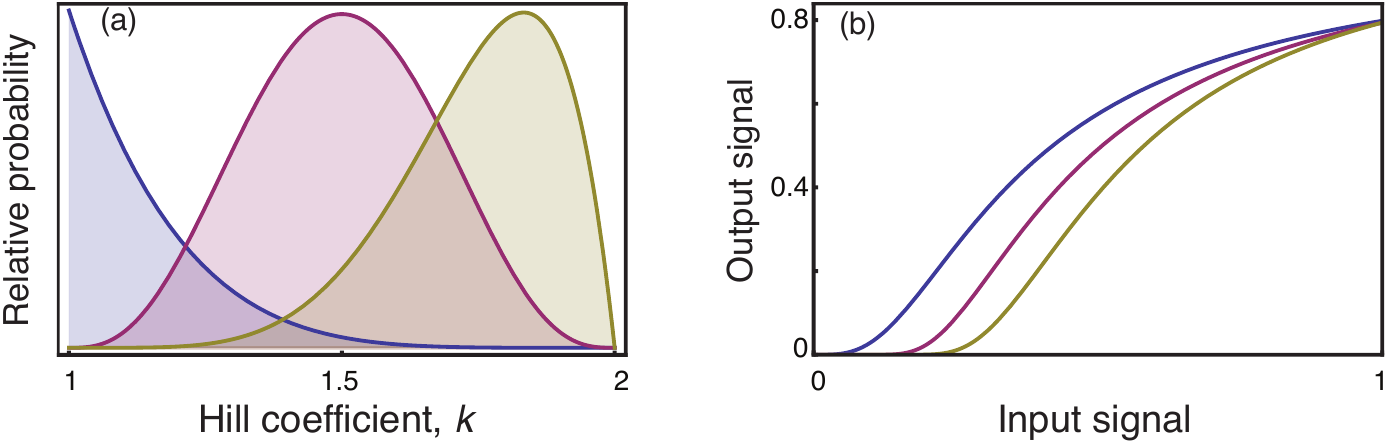}
\caption{\csentence{Signal processing cascade increases the Hill coefficient.}  The parameters for each reaction were chosen randomly from a beta distribution, denoted as a random variable $z\sim\mathrm{B}(\Ga,\Gb)$, which yields values in the range $[0,1]$.  The parameters $m=100z$ and $g=5+10z$ were chosen randomly and independently for each reaction from a beta distribution with $\Ga=\Gb=3$. The parameter $k$ for each reaction was obtained randomly as $1+z$, yielding a range of coefficients $1\le k\le 2$.  (a) In three separate trials, different combinations of $(\Ga,\Gb)$ were used for the beta distribution that generated the Hill coefficient, $k$: in the first, shown as the left distribution, $(\Ga,\Gb)=(1,6)$; in the second, shown in the middle, $(\Ga,\Gb)=(4,4)$; in the third, on the right, $(\Ga,\Gb)=(6,2)$.  The plot shows the peak heights normalized for each curve to be the same to aid visual comparison.  (b) The input-output relation over the full cascade. The curves from left to right correspond to the distributions for $k$ from left to right in the prior panel.  The input scale is normalized so that the maximum input value for each curve coincides at $80\%$ of the maximum output that could be obtained at infinite input.  The observed output curves have more strongly reduced sensitivity at low input than at high input compared with the Hill equation, but nonetheless match reasonably well.  The best fit Hill equation for the three curves has a Hill coefficient of, from left to right, $k=1.7, 2.2, 2.8$.  The average Hill coefficient for each reaction in a cascade is, from left to right, $k=1.14,1.5, 1.75$.  Each curve shows a single particular realization of the randomly chosen reaction parameters from the underlying distributions.  
\hbox{\null}
\label{fig:hillProd}
}
\end{figure}

A linear reaction cascade forms a simple type of system.  Kholodenko el al.\ \autocite{kholodenko97quantification} emphasized that a cascade tends to multiply the sensitivities of each step to determine the overall sensitivity of the system.  \Fig{hillProd} illustrates how the input-output relations of individual reactions combine to determine the system-level pattern.  

To generate \Fig{hillProd}, I calculated how a cascade of 12 reactions processes the initial input into the final output. Each reaction follows a Hill equation input-output relation given by \Eq{hill} with a half-maximal response at $m$ and a Hill coefficient of $k$. The output for each reaction is multiplied by a gain, $g$. The parameters for each reaction were chosen randomly, as shown in \Fig{hillProd}a and described in the caption.  

\Fig{hillProd} shows that a cascade significantly increases the Hill coefficient of the overall system above the average coefficient of each reaction, and often far above the maximum coefficient for any single reaction.  Intuitively, the key effect at low signal input arises because any reaction that has low sensitivity at low input reduces the signal intensity passed on, and such reductions at low input intensity multiply over the cascade, yielding very low sensitivity to low signal input.  Note in each curve that an input signal significantly above zero is needed to raise the output signal above zero.  That lower tail illustrates the loss of signal information at low signal intensity.

This analysis shows that weak logarithmic sensitivity at low signal input, associated with large Hill coefficients, can arise by aggregation of many reactions.  Thus, aggregation may be a partial solution to the overall puzzle of log-linear-log sensitivity in input-output relations.  

Aggregation by itself leaves open the question of how variations in sensitivity arise in the individual reactions.  Classical Michaelis-Menten reactions have linear sensitivity at low signal input, with a Hill coefficient of $k=1$.  A purely Michaelis-Menten cascade with $k=1$ at each step retains linear sensitivity at low signal input. A Michaelis-Menten cascade would not have the weak sensitivity at low input shown in \Fig{hillProd}b.  

How does a Hill coefficient $k>1$ arise in the individual steps of the cascade?  The power of aggregation to induce pattern means that it does not matter how such variations in sensitivity arise.  However, it is useful to consider some examples to gain of idea of the kinds of processes that may be involved beyond the deterministic cases given earlier.

\subsection*{Signal noise versus measurement noise}

Two different kinds of noise can influence the input-output relations of a system.  First, the initial input signal may be noisy, making it difficult for the system to discriminate between low input signals and background stochastic fluctuations in signal intensity \autocite{das09on-scaling}.  The classical signal-to-noise ratio problem expresses this difficulty by analyzing the ways in which background noise in the input can mask small changes in the average input intensity.  When the signal is weak relative to background noise, a system may be relatively insensitive to small increases in average input at low input intensity.

Second, for a given input intensity, the system may experience noise in the detection of the signal level or in the transmission of the signal through the internal processes that determine the final output.  Stochasticity in signal detection and transmission determine the measurement noise intrinsic to the system.  The ratio of measurement noise to signal intensity will often be greater at low signal input intensity, because there is relatively more noise in the detection and transmission of weak signals.  

In this section, I consider how signal noise and measurement noise influence Michaelis-Menten processes.  The issue concerns how much these types of noise may weaken sensitivity to low intensity signals. A weakening of sensitivity to low input distorts the input-output relation of a Michaelis-Menten process in a way that leads to a Hill equation type of response with $k>1$.  

In terms of measurement, Michaelis-Menten processes follow a linear-log scaling, in which sensitivity remains linear and highly precise at very low signal input intensity, and grades slowly into a logarithmic scaling with saturation.  By contrast, as the Hill coefficient, $k$, rises above one, measurement precision transforms into a log-linear-log scale, with weaker logarithmic sensitivity to signal changes at low input intensity.  Thus, the problem here concerns how signal noise or measurement noise may weaken input-output sensitivity at low input intensity. 

\subsubsection*{Input signal noise may not alter Michaelis-Menten sensitivity}

Consider the simplified Michaelis-Menten type of dynamics given in \Eq{mmKinetics}, repeated here for convenience
\begin{equation*}
  \dot{A} = gS(R-A) - \Gd A,
\end{equation*}
where $A$ is the output signal, $S$ is the input signal driving the reaction, $R$ is the reactant transformed by the input, $g$ is the rate of the transforming reaction which is a sort of signal gain level, and $\Gd$ is the rate at which the active signal output decays to the inactive reactant form.  Thus far, I have been analyzing this type of problem by assuming that the input signal, $S$, is a constant for any particular reaction, and then varying $S$ to analyze the input-output relation, given at equilibrium by Michaelis-Menten saturation
\begin{equation}\label{eq:mmEq2}
  A^* = g\left(\frac{S}{m+S}\right),
\end{equation}
where $m=\Gd/g$.  When input signal intensity is weak, such that $m\gg S$, then $A^*\approx gS$, which implies that output is linearly related to input.  

Suppose that $S$ is in fact a noisy input signal subject to random fluctuations.  How do the fluctuations affect the input-output relation for inputs of low average intensity?  Although the dynamics can filter noise in various ways, it often turns out that the linear input-output relation continues to hold such that, for low average input intensity, the average output is proportion to the average input, $\bar{A}\propto\bar{S}$. Thus, signal noise does not change the fact that the system effectively measures the average input intensity linearly and with essentially infinite precision, even at an extremely low signal to noise ratio.  

The high precision of classical chemical kinetics arises because the mass action assumption implies that infinitesimal changes in input concentration are instantly translated into a linear change in the rate of collisions between potential reactants.  The puzzle of Michaelis-Menten kinetics is that mass action implies high precision and linear scaling at low input intensity, whereas both intuition and observation suggest low precision and logarithmic scaling at low input intensity.  Input signal noise by itself typically does not alter the high precision and linear scaling of mass action kinetics.

Although the simplest Michaelis-Menten dynamics retain linearity and essentially infinite precision at low input, it remains unclear how the input-output relations of complex aggregate systems respond to the signal to noise ratio of the input.  Feedback loops and reaction cascades strongly influence the ways in which fluctuations are filtered between input and output.  However, classical analyses of signal processing tend to focus on the filtering properties of systems only in relation to fluctuations of input about a fixed mean value.  By contrast, the key biological problem is how input fluctuations alter the relation between the average input intensity and the average output intensity.  That key problem requires one to study the synergistic interactions between changes in average input and patterns of fluctuations about the average.  

For noisy input signals, what are the universal characteristics of system structure and signal processing that alter the relations between average input and average output?  That remains an open question.

\subsubsection*{Noise in signal detection and transmission reduces measurement precision and sensitivity at low signal input}

The previous section considered how stochastic fluctuations of inputs may affect the average output.  Simple mass action kinetics may lead to infinite precision at low input intensity with a linear scaling between average input and average output, independently of fluctuations in noisy inputs.  This section considers the problem of noise from a different perspective, in which the fluctuations arise internally to the system and alter measurement precision and signal transmission.

I illustrate the key issues with a simple model.  I assume that, in a reaction cascade with deterministic dynamics, each reaction leads to the Michaelis-Menten type of equilibrium input-output given in \Eq{mmEq2}. To study how stochastic fluctuations within the system affect input-output relations, I assume that each reaction has a certain probability of failing to transmit its input.  In other words, for each reaction, the output follows the equilibrium input-output relation with probability $1-p$, and with probability $p$, the output is zero. 

From the standard equilibrium in \Eq{mmEq2}, we simplify the notation by using $y\equiv A^*$ for output, and scale the input such that $x=S/m$.  The probability that the output is not zero is $1-p$, thus the expected output is
\begin{equation}\label{eq:zeroRx}
  y = g\left(\frac{x}{1+x}\right)(1-p).
\end{equation}
Let the probability of failure be $p=ae^{-bx}$. Note that as input signal intensity, $x$, rises, the probability of failure declines.  As the signal becomes very small, the probability of reaction failure approaches $a$, from the range $0\le a\le 1$.  

\begin{figure}[t]
\centering
\includegraphics[width=\hsize]{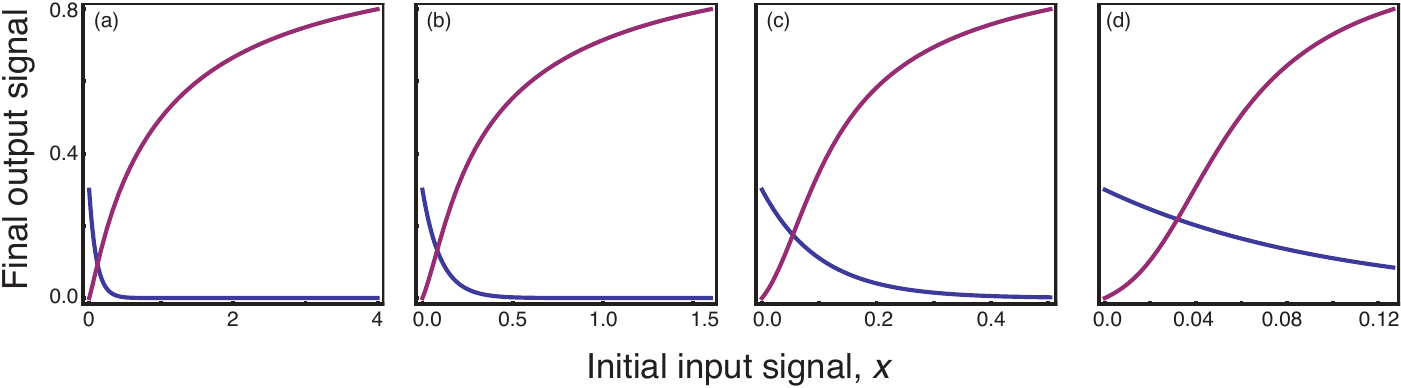}
\caption{\csentence{Stochastic failure of signal transmission reduces the relative sensitivity to low intensity input signals.}  The lower (blue) lines show the probability $p=ae^{-bx}$ that an input signal fails to produce an output. The upper (red) lines show the expected equilibrium output for Michaelis-Menten type dynamics corrected for a probability $p$ that the output is zero.  Each panel (a--d) shows a cascade of $n$ reactions, in which the output of each reaction forms the input for the next reaction, given an initial signal input of $x$ for the first reaction.  Each reaction follows \Eq{zeroRx}.  The number of reactions in the cascade increases from the left to the right panel as $n=1,2,4,8$. The other parameters for \Eq{zeroRx} are the gain per reaction, $g=1.5$, the maximum probability of reaction failure as the input declines to very low intensity, $a=0.3$, and the rate at which increasing signal intensity reduces reaction failure, $b=10$.  The final output signal is normalized to $0.8$ of the maximum output produced as the input become very large.
\hbox{\null}
\label{fig:signalLoss}
}
\end{figure}

\begin{figure}[t]
\centering
\includegraphics[width=\hsize]{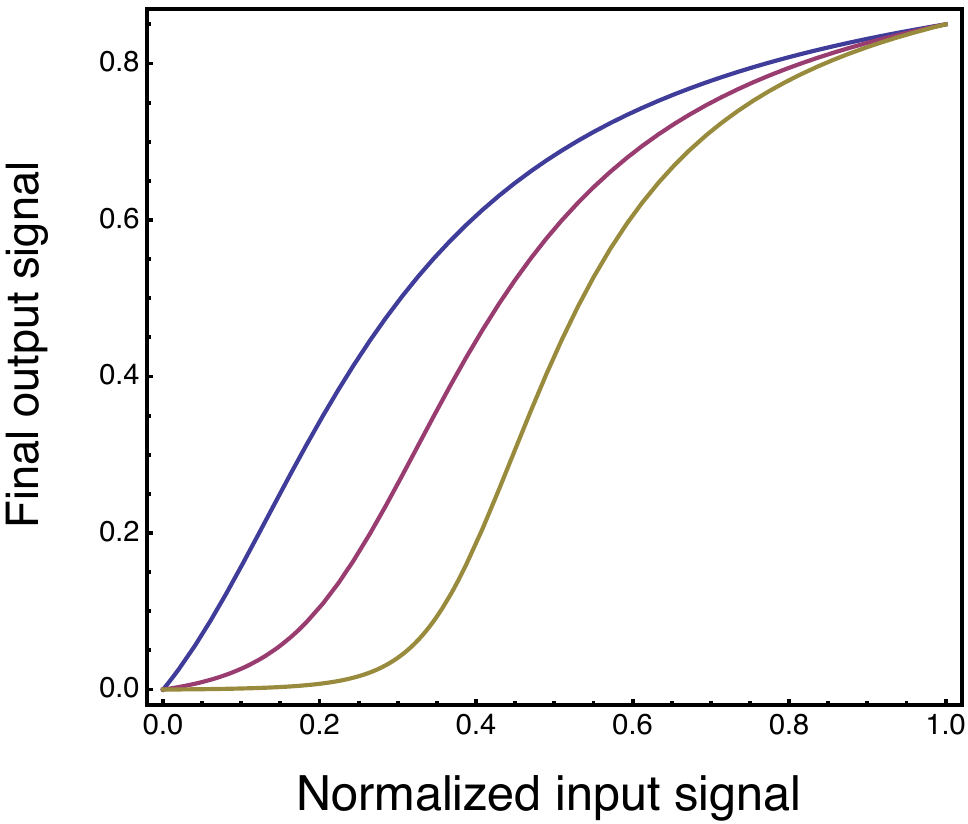}
\caption{\csentence{Greater failure rates for reactions reduce sensitivity to low input and increase the Hill coefficient, $k$.}   The curves arise from the same analysis as \Fig{signalLoss}, in which the curves from left to right are associated with an increase in the maximum failure rate as $a=0.2,0.4,0.6$. The curves in this figure have $n=8$ reactions in the cascade, a gain of $g=1.5$, and a decline in failure with increasing input, $b=10$. The scale for the input signal is normalized so that each curve has a final output of $0.85$ at a normalized input of one.  
\hbox{\null}
\label{fig:sigLossMax}
}
\end{figure}

\Fig{signalLoss} shows that stochastic failure of signal transmission reduces relative sensitivity to low input signals when a signal is passed through a reaction cascade.  The longer the cascade of reactions, the more the overall input-output relation follows an approximate log-linear-log pattern with an increasing Hill coefficient, $k$.  Similarly, \Fig{sigLossMax} shows that an increasing failure rate per reaction reduces sensitivity to low input signals and makes the overall input-output relation more switch-like.  

\subsubsection*{Implications for system design}

An input-output response with a high Hill coefficient, $k$, leads to switch-like function (\Fig{logIn}). By contrast, classical Michaelis-Menten kinetics lead to $k=1$, in which output increases linearly with small changes in weak input signals---effectively the opposite of a switch.  Many analyses of system design focus on this distinction.  The argument is that switch-like function will often be a favored feature of design, allowing a system to change sharply between states in response to external changes \autocite{tyson03sniffers}.  Because the intrinsic dynamics of chemistry are thought not to have a switch like function, the classical puzzle is how system design overcomes chemical kinetics to achieve switching function.

This section on stochastic signal failure presents an alternative view.  Sloppy components with a tendency to fail often lead to switch-like function.  Thus, when switching behavior is a favored phenotype, it may be sufficient to use a haphazardly constructed pathway of signal transmission coupled with weakly regulated reactions in each step.  Switching, rather than being a highly designed feature that demands a specific mechanistic explanation, may instead be the likely outcome of erratic biological signal processing.  

This tendency for aggregate systems to have a switching pattern does not mean that natural selection has no role and that system design is random.  Instead, the correct view may be that aggregate signal processing and inherent stochasticity set the contours of variation on which natural selection and system design work.  In particular, the key design features may have to do with modulating the degree of sloppiness or stochasticity. The distribution of gain coefficients in each reaction and the overall pattern of stochasticity in the aggregate may also be key loci of design.  

My argument is that systems may be highly designed, but the nature of that design can only be understood within the context of the natural patterns of variation.  The intrinsic contours of variation are the heart of the matter. I will discuss that issue again later.  For now, I will continue to explore the processes that influence the nature of variation in system input-output patterns.

\subsection*{Spatial correlations and departures from mass action}

Chemical reactions require molecules to come near each other spatially.  The overall reaction depends on the processes that determine spatial proximity and the processes that determine reaction rate given spatial proximity.  Roughly speaking, we can think of the spatial aspects in terms of movement or diffusion, and the transformation given spatial proximity in terms of a reaction coefficient.

Classical chemical kinetics typically assumes that diffusion rates are relatively high, so that spatial proximity of molecules depends only on the average concentration over distances much greater than the proximity required for reaction.  Kinetics are therefore limited by reaction rate given spatial proximity rather than by diffusion rate.  In contrast with classical chemical kinetics, much evidence suggests that biological molecules often diffuse relatively slowly, causing biological reactions sometimes to be diffusion limited (\Table{hill}).  

In this section, I discuss how diffusion-limited reactions can increase the Hill coefficient of chemical reactions, $k>1$.  That conclusion means that the inevitable limitations on the movement of biological molecules may be sufficient to explain the observed patterns of sensitivity in input-output functions and departure from Michaelis-Menten patterns.  

Two key points emerge. First, limited diffusion tends to cause potential reactants to become more spatially separated than expected under high diffusion and random spatial distribution. The negative spatial association between reactants arises because those potential reactants near each tend to react, leaving the nearby spatial neighborhood with fewer potential reactants than expected under spatial uniformity.  Negative spatial association of reactants reduces the rate of chemical transformation.  

This reduction in transformation rate is stronger at low concentration, because low concentration is associated with a greater average spatial separation of reactants.  Thus, low signal input may lead to relatively strong reductions in transformation rate caused by limited diffusion.  As signal intensity and concentration rise, this spatial effect is reduced.  The net consequence is a low transformation rate at low input, with rising transformation rate as input intensity increases.  This process leads to the the pattern characterized by higher Hill coefficients and switch-like function, in which there is low sensitivity to input at low signal intensity.

Limited diffusion within the broader context of input-output patterns leads to the second key point. I will suggest that limited diffusion is simply another way in which systems suffer reduced measurement precision and loss of information at low signal intensity.  The ultimate understanding of system design and input-output function follows from understanding how to relate particular mechanisms, such as diffusion or random signal loss, to the broader problems of measurement and information. To understand those broader and more abstract concepts of measurement and information, it is necessary to work through some of the particular details by which diffusion limitation leads to loss of information.

\subsubsection*{Departure from mass action}

Most analyses of chemical kinetics assume mass action.  Suppose, for example, that two molecules may combine to produce a bound complex
\begin{equation*}
	\cee{A + B ->[r] AB}
\end{equation*}
in which the bound complex, AB, may undergo further transformation.  Mass action assumes that the rate at which AB forms is $rAB$, which is the product of the concentrations of $A$ and $B$ multiplied by a binding coefficient, $r$.  The idea is that the number of collisions and potential binding reactions between A and B per unit of time changes linearly with the concentration of each reactant.  

Each individual reaction happens at a particular location.  That particular reaction perturbs the spatial association between reactants.  Those reactants that were, by chance, near each other, no longer exist as free potential reactants.  Thus, a reaction reduces the probability of finding potential reactants nearby, inducing a negative spatial association between potential reactants. To retain the mass action rate, diffusion must happen sufficiently fast to break down the spatial association. Fast diffusion recreates the mutually uniform spatial concentrations of the reactants required for mass action to hold.  

If diffusion is sufficiently slow, the negative spatial association between reactants tends to increase over time as the reaction proceeds.  That decrease in the proximity of potential reactants reduces the overall reaction rate.  Diffusion-limited reactions therefore have a tendency for the reaction rate to decline below the expected mass action rate as the reaction proceeds.  

That classical description of diffusion-limited reactions emphasizes the pattern of reaction rates over time.  By contrast, my focus is on the relation between input and output.  It seems plausible that diffusion limitation could affect the input-output pattern of a biological system.  But exactly how should we connect the classical analysis of diffusion limitation for the reaction rate of simple isolated reactions to the overall input-output pattern of biological systems?  

The connection between diffusion and system input-output patterns has received relatively little attention.  A few isolated studies have analyzed the ways in which diffusion limitation tends to increase the Hill coefficient, supporting my main line of discussion (\Table{hill}).  However, the broad field of biochemical and cellular responses has almost entirely ignored this issue.  The following sections present a simple illustration of how diffusion limitation may influence input-output patterns, and how that effect fits into the broader context of the subject.  

\subsubsection*{Example of input-output pattern under limited diffusion}

Limited diffusion causes spatial associations between reactants.  Spatial associations invalidate mass action assumptions.  To calculate reaction kinetics without mass action, one must account for spatially varying concentrations of reactants and the related spatial variations in chemical transformations.  There is no simple and general way to make spatially explicit calculations.  In some cases, simple approximations give a rough idea of outcome (\Table{hill}).  However, in most cases, one must study reaction kinetics by spatially explicit computer simulations.  Such simulations keep track of the spatial location of each molecule, the rate at which nearby molecules react, the spatial location of the reaction products, and the stochastic movement of each molecule by diffusion. 

Many computer packages have been developed to aid stochastic simulation of spatially explicit biochemical dynamics. I used the package Smoldyn \autocite{andrews04stochastic,andrews10detailed}.  I focused on the ways in which limited diffusion may increase Hill coefficients.  Under classical assumptions about chemical kinetics, diffusion rates tend to be sufficiently high to maintain spatial uniformity, leading to Michaelis-Menten kinetics with a Hill coefficient of $k=1$.  With lower diffusion rates, spatial associations arise, invalidating mass action.  Could such spatial associations lead to increased Hill coefficients of $k>1$?  

\begin{figure}[t!]
\centering
\includegraphics[width=\hsize]{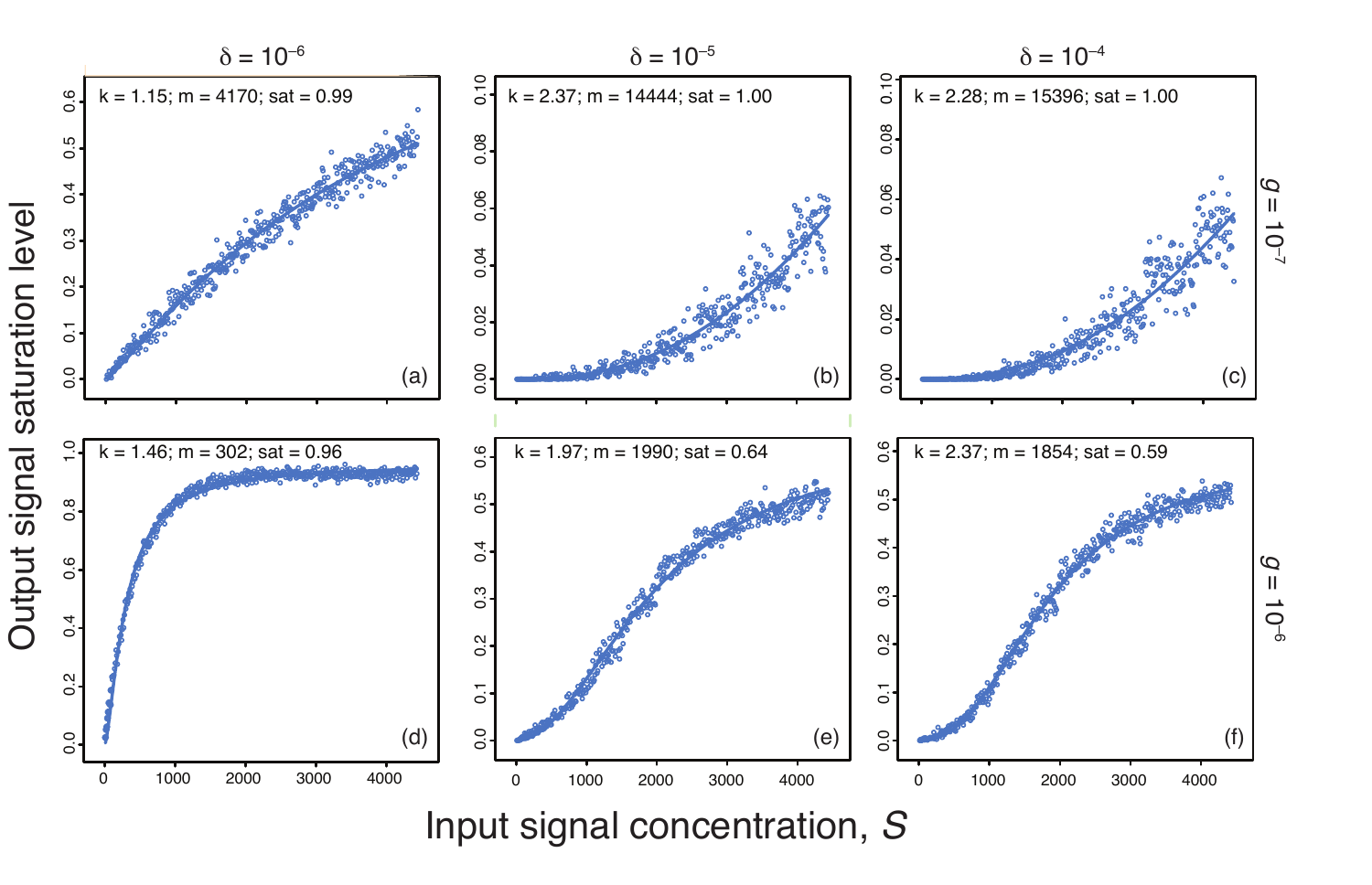}
\caption{\csentence{Limited diffusion and spatial association of reactants can increase the Hill coefficient, $k$.}  Simulations shown from the computer package Smoldyn, based on the reaction scheme in Eqs.~(\ref{eq:smoldReacta},\ref{eq:smoldReactb}). The concentration of the input signal, $S$, is the number of molecules per unit volume.  The other concentrations are set to $N=X=100$.  Diffusion rates are $10^{-5}$ for all molecules. I ran three replicates for each input concentration, $S$. Each circle shows the average of the three replicates. For each panel (a--f), I fit a Hill equation curve to the observations, denoting the output as the relative saturation level, $A/N = \mathrm{sat}\left[S^k\big/(m^k+S^k)\right]$. The fitted parameters are: $k$, the Hill coefficient; $m$, the input signal concentration that yields one-half of maximum saturation; and ``sat'', the maximum saturation level at which the output is estimated to approach an asymptotic value relative to the maximum theoretical value of one, at which all N has been transformed into A. Because of limited diffusion, actual saturation can be below the theoretical maximum of one.  Panels (b) and (c) are limited to output responses far below the median, because the simulations take too long to run for higher input concentrations.
\hbox{\null}
\label{fig:smoldyn}
}
\end{figure}

\Fig{smoldyn} shows clearly that increased Hill coefficients arise readily in a simple reaction scheme with limited diffusion. The particular reaction system is
\begin{align}
	\cee{S + R &->[g] S + A} \label{eq:smoldReacta}\\
	\cee{X + A &->[\Gd] X + R}. \label{eq:smoldReactb}
\end{align}
Under mass action assumptions, the dynamics would be identical to \Eq{mmKinetics}
\begin{equation*}
  \dot{A} = gS(N-A) - \Gd XA,
\end{equation*}
in which $N=R+A$ is the total concentration of inactive plus active reactant molecules and, in this case, we write the back reaction rate as $\Gd X$ rather than just $\Gd$ as in the earlier equation.  In a spatially explicit model, we must keep track of the actual spatial location of each X molecule, thus we need to include explicitly the concentration $X$ rather than include that concentration in a combined rate parameter.  At equilibrium, the output signal intensity under mass action follows the Michaelis-Menten relation
\begin{equation*}
  A^* = N\left(\frac{S}{m+S}\right),
\end{equation*}
in which $m=\Gd X/g$.  If we let $x=S/m$ and $y=A^*/N$, then we see that the reaction scheme here leads to an equilibrium input-output relation as in \Eq{hill} that follows the Hill equation
\begin{equation*}
  y = \left(\frac{x^k}{1+x^k}\right),
\end{equation*}
with $k=1$.  

I used the Smoldyn simulation package to study reaction dynamics when the mass action assumption does not hold. The simulations for this particular reaction scheme show input-output relations with $k>1$ when the rates of chemical transformation are limited by diffusion. \Fig{smoldyn} summarizes some Smoldyn computer simulations showing $k$ significantly greater than one for certain parameter combinations.  I will not go into great detail about these computer simulations, which can be rather complicated.  Instead, I will briefly summarize a few key points, because my goal here is simply to illustrate that limited diffusion can increase Hill coefficients under some reasonable conditions.

It is clear from \Fig{smoldyn} that limited diffusion can raise the Hill coefficient significantly above one.  What causes the rise?  It must be some aspect of spatial process, because diffusion limitation primarily causes departure from mass action by violating the assumption of spatial uniformity.  I am not certain which aspects of spatial process caused the departures in \Fig{smoldyn}.  It appeared that, in certain cases, most of the transformed output molecules, A, were maintained in miniature reaction centers, which spontaneously formed and decayed.  

A local reaction center arose when S and R molecules came near each other, transforming into S and A.  If there was also a nearby X molecule, then X and A caused a reversion to X and R. The R molecule could react again with the original nearby S molecule, which had not moved much because of a slow diffusion rate relative to the timescale of reaction.  The cycle could then repeat. If formation of reaction centers rises nonlinearly with signal concentration, then a Hill coefficient $k>1$ would follow.

Other spatial processes probably also had important, perhaps dominant, roles, but the miniature reaction centers were the easiest to notice.  In any case, the spatial fluctuations in concentration caused a significant increase in the Hill coefficient, $k$, for certain parameter combinations. 

\subsubsection*{Limited diffusion, measurement precision and information}

Why do departures from spatial uniformity and mass action sometimes increase the Hill coefficient?  Roughly speaking, one may think of the inactive reactant, R, as a device to measure the signal input concentration, $S$.  The rate of SR binding is the informative measurement.  The measurement scale is linear under spatial uniformity and mass action. The measurement precision is essentially perfect, because SR complexes form at a rate exactly linearly related to $S$, no matter how low the concentration $S$ may be and for any concentration $R$.

Put another way, mass action implies infinite linear measurement precision, even at the tiniest signal intensities.  By contrast, with limited diffusion and spatial fluctuations in concentration, measurement precision changes with the scale of the input signal intensity. For example, imagine a low concentration input signal, with only a few molecules in a local volume.  An SR binding transforms R into A, reducing the local measurement capacity, because it is the R molecules that provide measurement.  

With slow diffusion, each measurement alters the immediate capacity for further measurement.  The increase in information from measurement is partly offset by the loss in measurement capacity.  Put another way, the spatial disparity in the concentration of the measuring device R is a loss of entropy, which is a sort of gain in unrealized potential information.  As unrealized potential information builds in the spatial disparity of R, the capacity for measurement and the accumulation of information about S declines, perhaps reflecting a conservation principle for total information or, equivalently, for total entropy at steady state.

At low signal concentration, each measurement reaction significantly alters the spatial distribution of molecules and the measurement capacity. As signal concentration rises, individual reactions have less overall effect on spatial disparity.  Put another way, the spatial disparities increase as signal intensity declines, causing measurement to depend on scale in a manner that often leads to a logarithmic scaling.  I return to the problem of logarithmic scaling below.

\subsection*{Shaping sensitivity and dynamic range}

The previous sections considered specific mechanisms that may alter sensitivity of input-output relations in ways that lead to the log-linear-log scaling of the Hill equation.  Such mechanisms include stochastic failure of signal processing in a cascade or departures from mass action.  Those mechanisms may be important in many cases.  However, my main argument emphasizes that the widespread occurrence of log-linear-log scaling for input-output relations must transcend any particular mechanism.  Instead, general properties of system architecture, measurement and information flow most likely explain the simple regularity of input-output relations.  Those general properties, which operate at the system level, tend to smooth out the inevitable departures from regularity that must occur at smaller scales.

\subsubsection*{Brief review and setup of the general problem} 

An increase in the Hill coefficient, $k$, reduces sensitivity at low and high input signal intensity (\Fig{hillEx}).  At those intensities, small changes in input cause little change in output.  Weak sensitivity tends to be logarithmic, in the sense that output changes logarithmically with input.  Logarithmic sensitivities at low and high input often cause sensitivity to be strong and nearly linear within an intermediate signal range, with a rapid rate of change in output with respect to small changes in input intensity.  The intermediate interval over which high sensitivity occurs is the dynamic range.  The Hill coefficient often provides a good summary of the input-output pattern and is therefore a useful method for studying sensitivity and dynamic range.

The general problem of understanding biological input-output systems can be described by a simple question.  What processes shape the patterns of sensitivity and dynamic range in biological systems?  To analyze sensitivity and dynamic range, we must consider the architecture by which biological systems transform inputs to outputs.  

\subsubsection*{Aggregation of multiple transformations}

Biological systems typically process input signals through numerous transformations before producing an output signal.  Thus, the overall input-output pattern arises from the aggregate of the individual transformations.  Although the meaning of ``output signal'' depends on context, meaningful outputs typically arise from multiple transformations of the original input. 

I analyzed a simple linear cascade of transformations in an earlier section.  In that case, the first step in the cascade transforms the original input to an output, which in turn forms the input for the next step, and so on.  If individual transformations in the cascade have Hill coefficients $k>1$, the cascade tends to amplify the aggregate coefficient for the overall input-output pattern of the system.  Amplification occurs because weak logarithmic sensitivities at low and high inputs tend to multiply through the cascade.  Multiplication of logarithmic sensitivities at the outer ranges of the signal raises the overall Hill coefficient, narrows the dynamic range, and leads to high sensitivity over intermediate inputs.

That amplification of Hill coefficients in cascades leads back to the puzzle I have emphasized throughout this article.  For simple chemical reactions, kinetics follow the Michaelis-Menten pattern with a Hill coefficient of $k=1$.  If classical kinetics are typical, then aggregate input-output relations should also have Hill coefficients near to one.  By contrast, most observed input-output patterns have higher Hill coefficients.  Thus, some aspect of the internal processing steps must depart from classical Michaelis-Menten kinetics.

There is a long history of study with regard to the mechanisms that lead individual chemical reactions to have increased Hill coefficients.  In the first part of this article, I summarized three commonly cited mechanisms of chemical kinetics that could raise the Hill coefficient for individual reactions: cooperative binding, titration of a repressor, and opposing saturated forward and back reactions.  Those sorts of deterministic mechanisms of chemical kinetics do raise Hill coefficients and probably occur in many cases. However, the generality of raised Hill coefficients seems to be too broad to be explained by such specific deterministic mechanisms.  

\subsubsection*{Component failure}

If the classical deterministic mechanisms of chemical kinetics do not sufficiently explain the generality of raised Hill coefficients, then what does explain that generality?  My main argument is that input-output relations reflect underlying processes of measurement and information.  The nature of measurement and information leads almost inevitably to the log-linear-log pattern of observed input-output relations.  That argument is, however, rather abstract.  How do we connect the abstractions of measurement and information to the actual chemical processes by which biological systems transform inputs to outputs?

To develop the connection between abstract concepts and underlying mechanisms of chemical kinetics, I presented a series of examples.  I have already discussed aggregation, perhaps the most powerful and important general concept.  I showed that aggregation amplifies small departures from Michaelis-Menten kinetics $(k=1)$ into strongly log-linear-log patterns with increased $k$.  

In my next step, I showed that when individual components of an aggregate system have Michaelis-Menten kinetics but also randomly fail to transmit signals with a certain probability, the system converges on an input-output pattern with a raised Hill coefficient.  The main assumption is that failure rate increases as signal input intensity falls.  

Certainly, some reactions in biological systems will tend to fail occasionally, and some of those failures will be correlated with input intensity.  Thus, a small and inevitable amount of sloppiness in component performance of an aggregate system alters the nature of input-output measurement and information transmission.  Because the consequence of failures tends to multiply through a cascade, logarithmic sensitivity at low signal input intensity follows inevitably.  

Rather than invoke a few specific chemical mechanisms to explain the universality of log-linear-log scaling, this view invokes the universality of aggregate processing and occasional component failures.  I am not saying that component failures are necessarily the primary cause of log-linear-log scaling.  Rather, I am pointing out that such universal aspects must be common and lead inevitably to certain patterns of measurement and information processing.  Once one begins to view the problem in this way, other aspects begin to fall into place.

\subsubsection*{Departure from mass action}

Limited rates of chemical diffusion often occur in biological systems. I showed that limited diffusion may distort classical Michaelis-Menten kinetics to raise the Hill coefficient above one.  The increased Hill coefficient, and associated logarithmic sensitivity at low input, may be interpreted as reduced measurement precision for weak signals.

\subsubsection*{Regular pattern from highly disordered mechanisms}

The overall conclusion is that many different mechanisms lead to the same log-linear-log scaling.  In any particular case, the pattern may be shaped by the classical mechanisms of binding cooperativity, repressor titration, or opposing forward and back reactions.  Or the pattern may arise from the generic processes of aggregation, component failure, or departures from mass action.  

No particular mechanism necessarily associates with log-linear-log scaling.  Rather, a broader view of the relations between pattern and process may help. That broader view emphasizes the underlying aspects of measurement and information common to all mechanisms.  The common tendency for input-output to follow log-linear-log scaling may arise from the fact that so many different processes have the same consequences for measurement, scaling and information.  

The common patterns of nature are exactly those patterns consistent with the widest, most disparate range of particular mechanisms.  When great underlying disorder has, in the aggregate, a rigid common outcome, then that outcome will be widely observed, as if the outcome were a deterministic inevitability of some single underlying cause. The true underlying cause arises from generic aspects of measurement and information, not with specific chemical mechanisms.  

\subsubsection*{System design}

The inevitability of log-linear-log scaling from diverse underlying mechanisms suggests that the overall shape of biological input-output relations may be strongly constrained.  Put another way, the range of variation is limited by the tendency to converge to log-linear-log scaling.  However, within that broad class of scaling, biological systems can tune the responses in many different ways.  The tuning may arise by adjusting the number of reactions in a cascade, by allowing component failure rates to increase, by using reactions significantly limited by diffusion rate, and so on.  

Understanding the design of input-output relations must focus on those sorts of tunings within the broader scope of measurement and information transmission.  The demonstration that a particular mechanism occurs in a particular system is always interesting and always limited in consequence.  The locus of design and function is not the particular mechanism of a particular reaction, but the aggregate properties that arise through the many mechanisms that influence the tuning of the system.    

\subsubsection*{Robustness}

Overall input-output pattern often reflects the tight order that arises from underlying disorder. Thus, perturbations of particular mechanisms in the system may often have relatively little consequence for overall system function.  That insensitivity to perturbation---or robustness---arises naturally from the structure of signal processing in biological systems.  

To study robustness, it may not be sufficient to search for particular mechanisms that reduce sensitivity to perturbation.  Rather, one must understand the aggregate nature of variation and function, and how that aggregate nature shapes the inherent tendency toward insensitivity in systems \autocite{kauffman93the-origins,bluthgen03how-robust,frank09the-common}.  Once one understands the intrinsic properties of biological systems, then one can ask how those intrinsic properties are tuned by natural selection.

\subsection*{Measurement and information}

Intuitively, it makes sense to consider input-output relations with respect to measurement and information.  However, one may ask whether ``measurement'' and ``information'' are truly useful concepts or just vague and ultimately useless labels with respect to analyzing biological systems.  Here, I make the case that deep and useful concepts underlie ``measurement'' and ``information'' in ways that inform the study of biological design (\Table{psycho}).  I start by developing the abstract concepts in a more explicit way.  I then connect those abstractions to the nature of biological input-output relations.

\subsubsection*{Measurement}

Measurement is the assignment of a value to some underlying attribute or event.  Thus, we may think of input-output relations in biology as measurement relations. At first glance, this emphasis on measurement may seem trivial.  What do we gain by thinking of every chemical reaction, perception, or dose-response curve as a process of measurement?  

Measurement helps to explain why certain similarities in pattern continually arise.  When we observe common patterns, we are faced a question. Do common aspects of pattern between different systems arise from universal aspects of measurement or from particular mechanisms of chemistry or perception shared by different systems?  

Problems arise if we do not think about the distinction between general properties of measurement and specific mechanisms of particular chemical pathways.  If we do not think about that distinction, we may try to explain what is in fact a universal attribute of measurement by searching, in each particular system, for special aspects of chemical kinetics, pathway structure or physical laws that constrain perception.  In the opposite direction, we can never truly recognize the role of particular mechanisms in generating observed patterns if we do not separate out those aspects of pattern that arise from universal process.

Understanding universal aspects of pattern that arise from measurement means more than simply analyzing how observations are turned into numbers.  Instead, we must recognize that the structure of each problem sets very strong constraints on numerical pattern independently of particular chemical or biological mechanisms.  

\subsubsection*{Log-linear-log scales}

I have mentioned that the Hill equation is simply an expression of log-linear-log scaling.  The widely recognized value of the Hill equation for describing biological pattern arises from its connection to that underlying universal scale of measurement, in which small magnitudes scale logarithmically, intermediate magnitudes scale linearly, and large values scale logarithmically.  Although linear and logarithmic scales are widely used and very familiar, the actual properties and meanings of such scales are rarely discussed.  If we consider directly the nature of measurement scale, we can understand more deeply how to understand the relations between pattern and process.

Consider the example of measuring distance \autocite{frank10measurement}.  Start with a ruler that is about the length of your hand. With that ruler, you can measure the size of all the visible objects in your office. That scaling of objects in your office with the length of the ruler means that those objects have a natural linear scaling in relation to your ruler.

Now consider the distances from your office to various galaxies. Your ruler is of no use, because you cannot distinguish whether a particular galaxy moves farther away by one ruler unit. Instead, for two galaxies, you can measure the ratio of distances from your office to each galaxy. You might, for example, find that one galaxy is twice as far as another, or, in general, that a galaxy is some percentage farther away than another.
Percentage changes define a ratio scale of measure, which has natural units in logarithmic measure \autocite{hand04measurement}. For example, a doubling of distance always adds $\log(2)$ to the logarithm of the distance, no matter what the initial distance.

Measurement naturally grades from linear at local magnitudes to logarithmic at distant magnitudes when compared to some local reference scale. The transition between linear and logarithmic varies between problems. Measures from some phenomena remain primarily in the linear domain, such as measures of height and weight in humans. Measures for other phenomena remain primarily in the logarithmic domain, such as cosmological distances. Other phenomena scale between the linear and logarithmic domains, such as fluctuations in the price of financial assets \autocite{aparicio01empirical} or the distribution of income and wealth \autocite{dragulescu01exponential}.

Consider the opposite direction of scaling, from local magnitude to very small magnitude.  Your hand-length ruler is of no value for small magnitudes, because it cannot distinguish between a distance that is a fraction $10^{-4}$ of the ruler and a distance that is $2\times 10^{-4}$ of the ruler.  At small distances, one needs a standard unit of measure that is the same order of magnitude as the distinctions to be made.  A rule of length $10^{-4}$ distinguishes between $10^{-4}$ and $2\times 10^{-4}$, but does not distinguish between $10^{-8}$ and $2\times 10^{-8}$.  At small magnitudes, ratios can potentially be distinguished, causing the unit of informative measure to change with scale.  Thus, small magnitudes naturally have a logarithmic scaling. 

As we change from very small to intermediate to very large, the measurement scaling naturally grades from logarithmic to linear and then again to logarithmic, a log-linear-log scaling.  The locus of linearity and the meaning of very small and very large differ between problems, but the overall pattern of the scaling relations remains the same. This section analyzes that characteristic scaling in relation to the Hill equation and biological input-output patterns. I start by considering more carefully what measurement scales mean.  I then connect the abstract aspects of measurement to the particular aspects of the Hill equation and to examples of particular biological mechanisms.

\subsubsection*{Invariance, the essence of explanation}

We began with an observation. Many different input-output relations follow the Hill equation. We then asked: What process causes the Hill equation pattern?  It turned out that many very different kinds of process lead to the same log-linear-log pattern of the Hill equation.  We must change our question. What do the very different kinds of process have in common such that they generate the same overall pattern?

Consider two specific processes discussed earlier, cooperative binding and departures from mass action.  Those different processes may produce Hill equation patterns with similar Hill coefficients, $k$.  However, it is not immediately obvious why cooperative binding, departures from mass action, and many other different processes should lead to a very similar pattern. 

Group together all of the disparate mechanisms that generate a common Hill equation pattern.  When faced with a new mechanism, how can we tell if it belongs to the group?  We might look for particular features that are common to all members of the group.  However, that does not work.  Various potential members might have important common features. But the attributes that they do not share might cause one potential member to have a different pattern.  Common features are not sufficient.

More often common membership arises from the features that do not matter.  Think of circles. How can we describe whether a shape belongs to the circle class?  We have to say what does not matter. For circles, it does not matter how much one rotates them, they always look the same.  Circles are invariant to any rotation.  Equivalently, circles are symmetric with regard to any rotation.  Invariance and symmetry are the same thing.  Subject to some constraints, if a shape is invariant to any rotation, it is a circle.  If it is not invariant to all rotations, it is not a circle. The things that do not matter set the shared, invariant property of a group \autocite{feynman67the-character,anderson72more,weyl83symmetry}.  

A rotation is a kind of transformation.  The group is defined by the set of transformations that leave the group members unchanged, or invariant.  We can alter a chemical system from cooperative binding under mass action to noncooperative binding under departure from mass action, and the log-linear-log scaling may be preserved.  Such invariance arises because the different processes have an underlying symmetry with regard to the transformation of information from inputs to outputs (\Table{psycho}).  

What aspects of process do not matter with respect to causing the same log-linear-log pattern of the Hill equation?  How can we recognize the underlying invariance that joins together such disparate processes with respect to common pattern?  The Hill equation expresses measurement scale.  To answer our key questions, we must understand the meaning of measurement scale.  Measurement scale itself is solely an expression of invariance.  A particular measurement scale expresses what does not matter---the invariance under transformation that joins different kinds of processes to a common scaling.

\subsubsection*{Invariance and measurement}

Suppose a process transforms inputs $x$ to outputs $\GR(x)$.  The process may be a reading from a measurement instrument or a series of chemical transformations.  Given that process, how should we define the associated measurement scale?  Definitions can, of course, be made in any way.  But we should aim for something with reasonable meaning.  

One possible meaning for measurement is the scale that preserves information.  In particular, we seek a scale on which we obtain the same information from the values of the inputs as we do from the values of the outputs.  The measurement scale is the scale on which the input-output transformation does not alter the information in the signal  (\Table{psycho}).  

Information is, of course, often lost between input and output.  But only certain kinds of information are lost.  The measurement scale describes exactly what sort of information is lost during the transformation from input to output and what sort of information is retained.  In other words, the measurement scale defines the invariant qualities of information that remain unchanged by the input-output process.  

Different input-output processes belong to the same measurement scale when they share the same invariance that leaves particular aspects of information unchanged.  For such processes, certain aspects of information remain the same whether we have access to the original inputs or the final outputs when those values are given on the associated measurement scale.  By contrast, input-output processes that alter those same aspects of information when input and output values are given by a particular measurement scale do not belong to that scale.

Those abstract properties define a reasonable meaning for measurement scale. Such abstractness can be hard to parse.  However, it is essential to have a clear expression of those ideas, otherwise we could never understand why so many different kinds of biological processes can have such similar input-output relations, and why other processes do not share the same relations.  It is exactly those abstract informational aspects of measurement that unite cooperative binding and departures from mass action into a common group of processes that share a similar Hill equation pattern.

\subsubsection*{Measurement and information}

It is useful to express the general concepts in a simple equation.  I build up to that simple summary equation by starting with components of the overall concept.  

Inputs are given by $x$.  We denote a small change in input by $\dx$.  An input given on the measurement scale is $\Tr(x)$.  The sensitivity of the measurement scale to a change in input is
\begin{equation*}
	m_x = \dovr{\Tr(x)}{x},
\end{equation*}
which is the change on the measurement scale, $\dd\Tr(x)$, with respect to a change in input, $\dx$.  That sensitivity describes the information in the measurement scale with respect to fluctuations in inputs \autocite{frank10measurement,frank11a-simple,frank11measurement}. We may also write
\begin{equation*}
	m_x\dx = \dd\Tr(x),
\end{equation*}
providing an expression for the incremental information associated with a change in the underlying input, $\dx$.  If the scale is logarithmic, $\Tr(x)=\log(x)$, then 
\begin{equation*}
	m_x\dx = \dd\log(x)=\frac{\dx}{x},
\end{equation*}
for which the sensitivity of the measurement scale declines as the input becomes large.  On a purely logarithmic scale, the same increment in input, $\dx$, provides a lot of information when $x$ is small and little information when $x$ is large.  

Next, we express the relation that defines measurement scale.   On the proper measurement scale for a particular problem, the information from input values is proportional to the information from associated output values.  Put another way, the measurement scale is the transformation of values that makes information invariant to whether we use the input values or the output values.  The measurement scale reflects those aspects of information that are preserved in the input-output relation, and consequently also expresses those aspects of information that are lost in the input-output relation. Although rather abstract, it is useful to complete the mathematical development before turning to some examples in the next section. 

The output is $\GR(x)$, and the measurement scale transforms the output by $\Tr\left[\GR(x)\right]$. To have proportionality for the incremental information associated with a change in the underlying input, $\dd\Tr(x)$, and the incremental information associated with a change in the associated output, $\dd\Tr\left[\GR(x)\right]$, we have 
\begin{equation}\label{eq:measure}
	\dd\Tr(x) \propto \dd\Tr\left[\GR(x)\right]
\end{equation}
in which the $\propto$ relationship shows the proportionality of information associated with the sensitivity of inputs and outputs when expressed on the measurement scale.  That measurement scale defines the group of input-output processes, $\GR(x)$, that preserves the same invariant sensitivity and information properties on the scale $\Tr(x)$. In other words, all such input-output processes $\GR(x)$ that are invariant to the measurement scale transformation $\Tr(x)$ belong to that measurement scale \autocite{frank10measurement,frank11a-simple,frank11measurement}.  

In this equation, we have inputs, $x$, with the information in those inputs, $\dd\Tr(x)$, on the measurement scale $\Tr$, and outputs, $\GR(x)$, with information in those outputs, $\dd\Tr\left[\GR(x)\right]$, on the measurement scale $\Tr$.  We may abbreviate this key equation of measurement and information as 
\begin{equation*}
	\dd\Tr \propto \dd\Tr\left[\GR\right]
\end{equation*}
which we read as the information in inputs, $\dd\Tr$, is proportional to the information in outputs, $\dd\Tr\left[\GR\right]$.  All input-output relations $\GR(x)$ that satisfy this relation have the same invariant informational properties with respect to the measurement scale $\Tr$.  

\subsubsection*{Linear scale}

This view of measurement scale means that linearity has an exact definition.  Linearity requires that we obtain the same information from an increment $\dx$ on the input scale independently of whether the actual value is big or small (location), and whether we uniformly stretch or shrink all measurements by a constant amount.  To expresses changes in location and in uniform scaling, let
\begin{equation*}
	\Tr(x) = a + bx,
\end{equation*}
which changes the initial value, $x$, by altering the location by $a$ and the uniform stretching or shrinking (scaling) by $b$. This transformation is often called the linear transformation. But why is that the essence of linearity? From the first part of \Eq{measure}
\begin{equation*}
	m_x\dx = \dd\Tr(x) = b\,\dx \propto \dx,
\end{equation*}
which means that an increment in measurement provides a constant amount of information no matter what the measurement value, and that the information is uniform apart from a constant of proportionality $b$. Linearity means that information in measurements is independent of location and uniform scaling.  

What sort of input-output relations, $\GR(x)$, belong to the linear measurement scale? From the second part of \Eq{measure}, we have $\dd\Tr\left[\GR(x)\right]\propto \dx$, which we may expand as
\begin{align*}
  \dd\Tr\left[\GR(x)\right] &=\dd\left[a+b\GR(x)\right] \\
  		&= b\,\dd\GR(x) \propto \dx.
\end{align*}
Thus, any input-output relations such that $\dd\GR(x) \propto \dx$ belong to the linear scale, and any input-output relations that do not satisfy that condition do not belong to the linear scale.  To satisfy that condition, the input-output relation must have the form $\GR(x) = \Ga+\Gb x$, which is itself a linear transformation.  So, only linear input-output relations attach to a linear measurement scale.  If the input-output relation is not linear, then the proper measurement scale is not linear.

\subsubsection*{Logarithmic scale}

We can run the same procedure on the logarithmic measurement scale, for which a simple form is $\Tr(x)=\log(x)$.  For this scale, $\dd\Tr(x)=\dx/x$.  Thus, input-output relations belong to this logarithmic scale if 
\begin{align*}
  \dd\Tr\left[\GR(x)\right] &=\dd\log\left[\GR(x)\right]\\[6pt]
  		&= \frac{\dd\GR(x)}{\GR(x)} \propto \frac{\dx}{x}.
\end{align*}
This condition requires that $\GR(x)\propto x^k$, for which $\dd\GR(x)\propto x^{k-1}\dx$.  The logarithmic measurement scale applies only to input-output functions that have this power-law form (\Table{psycho}).  Note that the special case of $k=1$ leads to linear scaling, but for other $k$ values the scale is logarithmic.  

\subsubsection*{Linear-log and log-linear scales}

The most commonly used measurement scales are linear and logarithmic.  But those scales are unnatural, because the properties of measurement likely change with magnitude. As I mentioned earlier, an office ruler is fine for making linear measurements on the visible objects in your office.  But if you scale up to cosmological distances or down in microscopic distances, you naturally grade from linear to logarithmic.  A proper sense of measurement requires attention to the ways in which information and input-output relations change with magnitude \autocite{frank10measurement,frank11a-simple}.

Suppose an input increment provides information as
\begin{equation*}
	m_x\dx = \frac{\dx}{1+bx}.
\end{equation*}
When $x$ is small, $m_x\dx\approx\dx$, which is the linear measurement scale.  When $x$ is large, $m_x\dx\approx\dx/x$, which is the logarithmic scale.  The associated measurement scale is 
\begin{equation*}
  \Tr(x)\propto\log(1+bx),
\end{equation*} 
and the associated input-output functions satisfy $\GR(x)\propto (1+bx)^k$.  This scale grades continuously from linear to logarithmic.  The parameter $b$ determines the relation between magnitude and the type of scaling.   

The inverse scaling grades from logarithmic at small magnitudes to linear as magnitude increases, with 
\begin{equation*}
  \Tr(x)\propto x+b\log(x).
\end{equation*} 
When $x$ is small, the scale is logarithmic with $\Tr(x)\approx b\log(x)$. When $x$ is large, the scale is linear with $\Tr(x)\approx x$. 

\subsection*{Biological input-output: log-linear-log}

I have emphasized that the log-linear-log scale is perhaps the most natural of all scales.  Information in measurement increments tends to be logarithmic at small and large magnitudes.  As one moves in either extreme direction, the unit of measure changes in proportion to magnitude to preserve consistent information.  At intermediate magnitudes, changing values associate with an approximately linear measurement scale.  For many biological input-output relations, that intermediate, linear zone is roughly the dynamic range.  

The Hill equation description of input-output relations
\begin{equation*}
  \GR(x)=\frac{x^k}{1+x^k},
\end{equation*}
is widely useful because it describes log-linear-log scaling in a simple form.  To check for log scaling in the limits of high or low input, we use $\Tr(x)=\log(x)$, which implies $\dd\Tr(x)\propto\dd x/x$. In our fundamental relation of measurement, we have
\begin{align*}
  \dd\Tr(x)&\propto\dd\Tr\left[\GR(x)\right]\\
  	&=\dd\log\left[\GR(x)\right] \\
  	&=k\left(\frac{1}{x}-\frac{x^{k-1}}{1+x^k}\right)\dx,
\end{align*}
When $x$ is small, $\dd\Tr(x)\propto \dx/x$, the expression for input-output functions associated with the logarithmic scale.  When $x$ is large, $\dd\Tr(x)\propto -\dx/x$, which is the expression for saturation on a logarithmic scale.  

When $k>1$, the input-output relation scales linearly for intermediate $x$ values.  One can do various calculations to show the approximate linearity in the middle range.  But the main point can be seen by simply looking at \Fig{hillEx}. 

\begin{figure}[t!]
\centering
\includegraphics[width=\hsize]{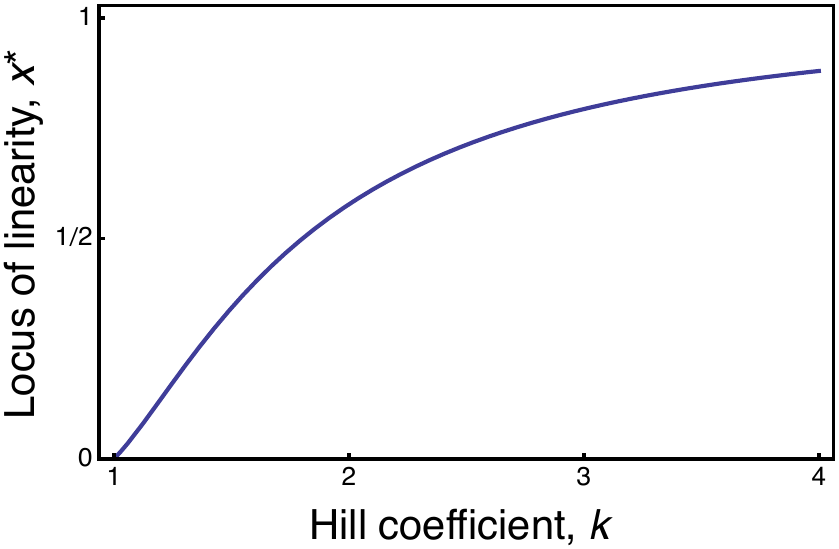}
\caption{\csentence{The locus of linearity, which is the value of input, $x^*$, at which the log-linear-log pattern of the Hill equation becomes exactly linear.}   The locus of linearity corresponds to the peak sensitivity of the input-output relation. At $x^*=1$, output is one-half of maximal response. Plot based on \Eq{linearity}.  
\hbox{\null}
\label{fig:linearity}
}
\end{figure}

Exact linearity occurs when the second derivative of the Hill equation vanishes at 
\begin{equation}\label{eq:linearity}
  x^*=\left(\frac{k-1}{k+1}\right)^{1/k}
\end{equation}
for $k>1$. \Fig{linearity} shows that the locus of linearity shifts from the low side as $k\rightarrow 1$ and $x^*\rightarrow 0$ to the high side as $k\rightarrow\infty$ and $x^*\rightarrow 1$. Note that $x^*=1$ is the input at which the response is one-half of the maximum.

\subsubsection*{Sensitivity and information}

Sensitivity is the responsiveness of output for a small change in input.  For a log-linear-log pattern, the locus of linearity is often equivalent to maximum sensitivity of the output in relation to the input.  The logarithmic regimes at low and high input are relatively weakly sensitive to changes in input.

The Hill equation pattern for input, $x$, and output, $\GR(x)$, is
\begin{equation*}
  \GR(x)=\frac{x^k}{1+x^k}=\frac{1}{1+e^{-k\log(x)}}.
\end{equation*}
The equivalent form on the right side is the classic logistic function expressed in terms of $\log(x)$ rather than $x$. This logarithmic form is the log-logistic function. Note also that $\GR(x)$ varies between zero and one as $x$ increases from zero.  Thus, $\GR(x)$ is analogous to a cumulative distribution function (cdf) from probability theory.  These mathematical analogies for input-output curves will be useful as we continue to analyze the meaning of input-output relations and why certain patterns are particularly common. 

Note also that $k=1$ is the Michaelis-Menten pattern of chemical kinetics.  This relation of the input-output curve $\GR(x)$ to chemical kinetics will be important when we connect general aspects of sensitivity to the puzzles of chemical kinetics and biochemical input-output patterns.

The sensitivity is the change in output with respect to input. Thus, sensitivity is the derivative of $\GR$ with respect to $x$, which is
\begin{equation*}
  \Gdot(x)=\frac{kx^{k-1}}{(1+x^k)^2}.
\end{equation*}
This expression is analogous to the log-logistic probability distribution function (pdf). Here, I obtained the pdf in the usual way by differentiating the cdf. Noting that the pdf is the sensitivity of the cdf to small changes in value (input), we have an analogy between the sensitivity of input-output relations and the general relation between the pdf and cdf of a probability distribution.

Maximum sensitivity is the maximum value of $\Gdot(x)$, which corresponds to the mode of the pdf.  For $k\le1$, the maximum occurs at $x=0$, which means that measurement sensitivity of the input-output system is greatest when the input is extremely small.  Intuitively, it seems unlikely that maximum sensitivity could be achieved when discriminating tiny input values.  For $k>1$, the maximum value of the log-logistic pattern occurs when $\Gddot(x)=0$, which is the point at which the second derivative is zero and the input-output relation is purely linear.  That maximum occurs at the point given in \Eq{linearity}.  

The analogy with probability provides a connection between input-output functions, measurement and information.  A probability distribution is completely described by the information that it expresses \autocite{jaynes03probability,frank09the-common}.  That information can be split into two parts. First, certain constraints must be met that limit the possible shapes of the distribution, such as the mean, the variance, and so on. Second, the measurement scale sets the sensitivity of the outputs in terms of randomness (entropy) and information (negative entropy) in relation to changes in observed values or inputs \autocite{frank10measurement,frank11a-simple}.  

\subsubsection*{Sensitivity, measurement and the shape of input-output patterns}

The Hill equation seems almost magical in its ability to fit the input-output patterns of diverse biological processes.  The magic arises from the fact that the Hill equation is a simple expression of log-linear-log scaling when the Hill coefficient is $k>1$.  The Hill coefficient expresses the locus of linearity.  As $k$ declines toward one, the pattern becomes linear-log, with linearity at low input values grading into logarithmic as input increases.  As $k$ drops below one, the pattern becomes everywhere logarithmic, with declining sensitivity as input increases.  

Sensitivity and measurement scale are the deeper underlying principles. The Hill equation is properly viewed as just a convenient mathematical form that expresses a particular pattern of sensitivity, measurement, and the informational properties of the input-output pattern.  From this perspective, one may ask whether alternative input-output functions provide similar or better ways to express the underlying log-linear-log scale.  

Frank \& Smith \autocite{frank10measurement,frank11a-simple} presented the general relations between measurement scales and associated probability distribution function (pdf) patterns.  Because a pdf is analogous to an expression of sensitivity for input-output functions, we can use their system as a basis for alternatives to the Hill equation.  Perhaps the most compelling general expressions for log-linear-log scales arise from the family of beta distributions. For example, the generalized beta prime distribution can be written as
\begin{equation}\label{eq:betaPrime}
	\Gdot(x) \propto \left(\frac{x}{m}\right)^{\Ga}\left(1+\left(\frac{x}{m}\right)^k\right)^{-\Gb}.
\end{equation} 
With $\Ga=k$ and $\Gb=1$, we obtain a typical form of the Hill equation given in \Eq{hilldim}.  The additional parameters $\Ga$ and $\Gb$ provide more flexibility in expressing different logarithmic sensitivities at high versus low inputs.  

The theory of measurement scale and probability in Frank \& Smith \autocite{frank10measurement,frank11a-simple} also provides a way to analyze more complex measurement and sensitivity schemes.  For example, a double log scale (logarithm of a logarithm) reduces sensitivity below classical single log scaling.  Such double log scales provide a way to express more extreme dissipation of signal information in a cascade at low or high input levels.  

These different expressions for sensitivity have two advantages.  First, they provide a broader set of empirical relations to use for fitting data.  Those empirical relations derive from the underlying principles of measurement scale.  Second, the different forms express hypotheses about how signal processing cascades dissipate information in signals and alter patterns of sensitivity.  For example, one may predict that certain signal cascade architectures dissipate information more strongly and lead to double logarithmic scaling and loss of sensitivity at certain input levels.  Further theory could help to sort out the predicted relations between signal processing architecture, the dissipation of information, and the general forms of input-output relations.

\section*{Conclusions}

Nearly all aspects of biology can be reduced to inputs and outputs.  A chemical reaction is the transformation of input concentrations to output concentrations. Developmental or regulatory subsystems arise from combinations of chemical reactions. Any sort of sensory measurement of environmental inputs follows from chemical output responses.  The response of a honey bee colony to changes in temperature or external danger follows from perceptions of external inputs and the consequent output responses.  Understanding biology mostly has to do with description of input-output patterns and understanding the processes that generate those patterns.

I focused on one simple pattern, in which outputs rise with increasing inputs.  I emphasized basic chemistry for two reasons.  First, essentially all complex biological processes reduce to cascades of simple chemical reactions.  Understanding complex systems ultimately comes down to understanding the relation between combinations of simple reactions and the resulting patterns at the system level.  Second, the chemical level presents a clear puzzle.  The classical theory of chemical kinetics predicts a concave Michaelis-Menten input-output relation.  By contrast, many simple chemical reactions follow an S-shaped Hill equation pattern.  The input-output relations of many complex systems also tend to follow the Hill equation.

I analyzed this distinction between Michaelis-Menten kinetics and Hill equation patterns in order to illustrate the broad problems posed by input-output relations. Several conclusions follow.

First, many distinct chemical processes lead to the Hill equation pattern.  The literature mostly considers those different processes as a listing of exceptions to the classical Michaelis-Menten pattern.  Each observed departure from Michaelis-Menten is treated as a special case requiring an explicit mechanistic explanation chosen from the list of possibilities. 

Second, I emphasized an alternative perspective. A common pattern is widespread because it is consistent with the greatest number of distinct underlying mechanisms.  Thus, the Hill equation pattern may be common because there are so many different processes that lead to that outcome.  

Third, because a particular common pattern associates with so many distinctive underlying processes, it is a mistake to treat each observed case of that pattern as demanding a match to a particular underlying process.  Rather, one must think about the problem differently.  What general properties cause the pattern to be common?  What is it about all of the different processes that lead to the same outcome?  

Fourth, I suggested that aggregation provides the proper framing.  Roughly speaking, aggregation concerns the structure by which different components combine to produce the overall input-output relations of the system.  The power of aggregation arises from the fact that great regularity of pattern often emerges from underlying disorder.  Deep understanding turns on the precise relation between underlying disorder and emergent order.

Fifth, measurement in relation to the dissipation of information sets the match between underlying disorder and emergent order.  The aggregate combinations of input-output processing that form the overall system pattern tend to lose information in particular ways during the multiple transformations of the initial input signal.  The remaining information carried from input to output arises from aspects of precision and measurement in each processing step.  

Sixth, previous work on information theory and probability shows how aggregation may influence the general form of input-output relations. In particular, certain common scaling relations tend to set the invariant information carried from inputs to outputs.  Those scaling relations and aspects of measurement precision tell us how to evaluate specific mechanisms with respect to their general properties.  Further work may allow us to classify apparently different processes into a few distinctive sets.

Seventh, classifying processes by their key properties may ultimately lead to a meaningful and predictive theory.  By that theory, we may understand why apparently different processes share similar outcomes, and why certain overall patterns are so common.  We may then predict how overall pattern may change in relation to the structural basis of aggregation in a system and the general properties of the underlying components.  More theoretical work and associated empirical tests must follow up on that conjecture. 

Eighth, I analyzed the example of fundamental chemical kinetics in detail.  My analysis supports the general points listed here.  Specific analyses of other input-output relations in terms of aggregation, measurement and scale will provide the basis for a more general theory.  

Ninth, robustness means insensitivity to perturbation.  Because system input-output patterns tend to arise by the regularities imposed by aggregation, systems naturally express order arising from underlying disorder in components.  The order reflects broad structural aspects of the system rather than tuning of particular components.  Perturbations to individual components will therefore tend to have relatively little effect on overall system performance---the essence of robustness.

Finally, natural selection and biological design may be strongly influenced by the regularity of input-output patterns.  That regularity arises inevitably from aggregation and the dissipation of information.  Those inevitably regular patterns set the contours that variation tends to follow.  Thus, biological design will also tend to follow those contours.  Natural selection may act primarily to modulate system properties within those broad constraints. How do changes in extrinsic selective pressures cause natural selection to alter overall system architecture in ways that modulate input-output patterns?


\begin{backmatter}

\section*{Competing interests}
  The author declares that he has no competing interests.

\section*{Acknowledgements}

I developed this work while supported by a Velux Foundation Professorship of Biodiversity at ETH Z\"urich.  I benefitted greatly from many long discussions with Paul Schmid-Hempel.

\section*{Grant information}

National Science Foundation (USA) grants EF--0822399 and DEB--1251035 support my research.  


\bibliographystyle{bmc-mathphys} 
\bibliography{main}      







%
%

\end{backmatter}
\end{document}